\documentclass[a4paper,fleqn,useAMS,usenatbib]{mnras}
\usepackage{newtxtext,newtxmath}
\usepackage[T1]{fontenc}
\usepackage{ae,aecompl}
\usepackage{epsfig}
\usepackage{amsmath, amsfonts, epsfig, xspace}
\usepackage{natbib}
\usepackage{rotating}
\usepackage{graphicx}
\usepackage{caption}
\usepackage{longtable}
\usepackage{multicol}
\usepackage{booktabs}
\usepackage{changepage}
\usepackage{placeins}
\usepackage{scalerel}

\def\betanir{$\beta_{\scaleto{\rm NIR \rm}{3.5pt}}$~}
\def\NSbetanir{$\beta_{\scaleto{\rm NIR \rm}{3.5pt}}$}

\def\fraccon{${\Delta \rm F}/{\langle \rm F \rangle_{1700}}$~}
\def\fracew{${\Delta \rm EW}/{\langle \rm EW \rangle}$~}

\def\delew{${\Delta \rm EW}$~}
\def\avgew{${\langle \rm EW \rangle}$~}
\def\deld{${\Delta \rm d_{\scaleto{\rm BAL \rm}{3.5pt}} }$~}
\def\delv{${\Delta \rm v}$~}
\def\avgd{${\langle \rm d_{\scaleto{\rm BAL \rm}{3.5pt}} \rangle}$~}
\def\delalp{${\Delta \rm \alpha}$~}
\def\delt{${\Delta \rm t}$~}
\def\delm{${\Delta \rm m}$~}
\def\meddelm{median (${\Delta \rm m}$)~}
\def\sigdelm{$\sigma_{\Delta \rm m}$~}
\def\slopedelm{${\Delta \rm m / \Delta \rm t}$~}

\def\NSfraccon{${\Delta \rm F}/{\langle \rm F \rangle_{1700}}$}
\def\NSfracew{${\Delta \rm EW}/{\langle \rm EW \rangle}$}
\def\NSfracewsig{$\sigma_{{\Delta \rm EW}/{\langle \rm EW \rangle}}$}
\def\NSdelew{${\Delta \rm EW}$}
\def\NSavgew{${\langle \rm EW \rangle}$}
\def\NSdeld{${\Delta \rm d_{\scaleto{\rm BAL \rm}{3.5pt}} }$}
\def\NSdelv{${\Delta \rm v}$}
\def\NSavgd{${\langle \rm d_{\scaleto{\rm BAL \rm}{3.5pt}} \rangle}$}
\def\NSdelalp{${\Delta \rm \alpha}$}
\def\NSdelt{${\Delta \rm t}$}
\def\NSdelm{${\Delta \rm m}$}
\def\NSmeddelm{median (${\Delta \rm m}$)}
\def\NSsigdelm{$\sigma_{\Delta \rm m}$}
\def\NSslopedelm{${\Delta \rm m / \Delta \rm t}$}

\def\maincomplex{{\it Preliminary}~}

\def\emii{{\it Appearing}~}
\def\emiipri{{\it Appearing pristine}~}
\def\dismain{{\it Disappearing}~}

\def\NSemii{{\it Appearing}}
\def\NSemiipri{{\it Appearing pristine}}
\def\NSdismain{{\it Disappearing}}

\def\pfour{P$_{4}$~}
\def\peight{P$_{8}$~}

\def\NSpfour{P$_{4}$}
\def\NSpeight{P$_{8}$}
\def\NSdispri{Pristine}

\def\civab{C~{\sc iv}$\lambda\lambda$1548,1550~}
\def\sivab{S~{\sc iv}$\lambda\lambda$1393,1403~}
\def\ciiia{C~{\sc iii]}$\lambda$1909~}

\def\mgii{Mg~{\sc ii}~} 
\def\feii{Fe~{\sc ii}~} 

\def\feii{Fe~{\sc ii}~}
\def\feiii{Fe~{\sc iii}~}
\def\aliii{Al~{\sc iii}~}

\def\pv{P~{\sc v}~}

\def\civ{C~{\sc iv}~}
\def\cv{C~{\sc v}~}
\def\ciii{C~{\sc iii]}~}
\def\siiv{Si~{\sc iv}~}
 
\def\NSmgii{Mg~{\sc ii}}

\def\NSaliii{Al~{\sc iii}}
\def\NSalii{Al~{\sc ii}}

\def\NSnv{N~{\sc v}}
\def\NSciv{C~{\sc iv}}

\def\NSsiiv{Si~{\sc iv}}

\def\zem{$z_{\rm em}$}

\def\kms{~km~s$^{-1}$~}
\def\NSkms{~km~s$^{-1}$}
\def\aj{{AJ}}%
\def\apj{{ApJ}}%
\def\apjl{{ApJ}}%
\def\apjs{{ApJS}}%
\def\aap{{A\&A}}%
\def\mnras{{MNRAS}}%
\def\pasp{{PASP}}%
\title[$Mishra$ et al.]{Appearance vs Disappearance of broad absorption line troughs in quasars}
\author[$Mishra$ et al.]{{\Large Sapna Mishra$^{1, 2}$\thanks{E-mail: sapna@aries.res.in(SM)}, M. Vivek$^{3}$, Hum Chand$^{4, 1}$, and Ravi Joshi$^{5}$}\\\\
  $^{1}$Aryabhatta Research Institute of Observational Sciences (ARIES), Manora Peak, Nainital $-$ 263002, India\\
  $^{2}$Department of Physics \& Astrophysics, University of Delhi, Delhi 110 007, India\\
  $^{3}$Indian Institute of Astrophysics, Koramangala, Bengaluru 560 034, India\\
  $^{4}$Department of Physics and Astronomical Sciences, Central University of Himachal Pradesh (CUHP), Dharamshala-176215, India\\
  $^{5}$Kavli Institute for Astronomy and Astrophysics, Peking University, Beijing 100871, China\\
}

\begin{document}
\date{Accepted ---. Received ---; in original form ---}

\pagerange{\pageref{firstpage}--\pageref{lastpage}} \pubyear{2019}

\maketitle

\label{firstpage}
\begin{abstract}
We present a new set of 84 Broad absorption line (BAL) quasars ( 1.7 $<$ \zem $<$ 4.4)  exhibiting an appearance of \civ BAL troughs over 0.3$-$4.8 rest-frame years by comparing the Sloan Digital Sky Survey Data Release (SDSSDR)-7, SDSSDR-12, and SDSSDR-14 quasar catalogs. We contrast the nature of BAL variability in this appearing BAL quasar sample with a disappearing BAL quasar sample studied in literature by comparing the quasar's intrinsic, BAL trough, and continuum parameters between the two samples. We find that appearing BAL quasars have relatively higher redshift and smaller probed timescales as compared to the disappearing BAL quasars. To mitigate the effect of any redshift bias, we created control samples of appearing and disappearing BAL quasars that have similar redshift distribution.  We find that the appearing BAL quasars are relatively brighter and have shallower and wider BAL troughs compared to the disappearing BAL sample. The distribution of quasar continuum variability parameters between the two samples is clearly separated, with the appearance of the BAL troughs being accompanied by the dimming of the continuum and vice versa. Spectral index variations in the two samples also point to the anti-correlation between the BAL trough and continuum variations consistent with the "bluer when brighter" trend in quasars. We show that the intrinsic dust model is less likely to be a favorable scenario in explaining BAL appearance/disappearance. Our analysis suggests that the extreme variations of BAL troughs like BAL appearance/disappearance are mainly driven by changes in the ionization conditions of the absorbing gas. 
\end{abstract}


\section{Introduction}
\label{sec:intro}

Broad absorption line (BAL) quasars are among the 10$-$15\% population of quasars exhibiting strong, broad absorption troughs blue-shifted relative to the quasar's emission redshift in the rest-frame ultra-violet (UV) spectra. The presence of broad absorption is thought to be arising from the co-existing optically thick material in the outflowing winds generated near the accretion disk and accelerated radially outward \citep{1995ApJ...451..498M,2000ApJ...545...63E}. These winds often carry a large amount of energy and momentum along their way and are hence a potential contributor to the active galactic nuclei (AGN) feedback \citep{doi:10.1146/annurev-astro-081811-125521}.\par

The conventional definition of BALs demands the absorption feature to be wider than 2000\kms and at least 10\% below the continuum level \citep{1991ApJ...373...23W}.  Additionally, absorption troughs with a width between 500$-$2000\kms are known as mini-BALs, while those with widths less than 500\kms are termed as narrow absorption lines (NALs). Further, BALs are divided into three sub-classes based on the ionization
states of absorption troughs. BAL troughs resulting from the high ionization transitions of the species such as \NSciv, \NSsiiv, \NSnv, \pv\ are referred to as HiBALs \citep{1991ApJ...373...23W}.  They contribute to the majority of the BAL population. Around $\sim$10\% of BALs arise from the transition of lower ionization species such as \NSalii, \NSaliii, \mgii and are known as LoBALs \citep{1993ApJ...413...95V}. Very few BAL populations show absorption troughs of the transitions from the excited states of \feii or \feiii and are called FeLoBALs \citep{2000ApJ...538...72B}.\par

Many BAL outflows are believed to be originated from the accretion disk at 10$-$100 light days from the super massive black hole (SMBH) \citep[e.g.][]{1995ApJ...451..498M,1998MNRAS.295..595P}, then the time required for outflowing clouds with a typical outflow velocity of 8000\kms to cross the launching region will be about 1$-$10 years. Therefore, it is believed that fluctuations produced in the accretion disk can give rise to the variations of BAL outflows over the timescales of 1$-$10 years.  Thus, multi-year variability studies of BAL are important to understand the location and physical conditions (e.g., quasar wind lifetimes, sizes, and geometries) of the absorbing gas and the physical mechanisms responsible for these outflows. Variations of BAL troughs such as changes in the absorption strength, e.g., equivalent width (EW), appearance or disappearance of BAL troughs, kinematic shifts in the absorption profiles, and changes in the shape of BAL profiles in \civ and \siiv BALs have been studied in several cases. Such variations in the BAL absorption profiles can be comprehended from (i) the changes in covering fraction of the quasar by the absorbing gas due to its transverse motions across our line of sight \citep{2007ApJ...656...73L,2008ApJ...675..985G,2011MNRAS.413..908C,2013MNRAS.429.1872C,2014MNRAS.440..799V,2019ApJ...870L..25Y}, and/or (ii) changes in the ionizing  radiation giving rise to  changes in optical depth of the absorbers \citep{1994PASP..106..548B,2012MNRAS.422.3249C,2015MNRAS.454.3962H,2018MNRAS.473L.106L,2018ApJ...862...22R,2019MNRAS.486.2379V}, (iii) changes in the acceleration profile and/or geometry of the outflow due to changes in the driving force \citep{2014MNRAS.442..862J,2016ApJ...824..130G,2019ApJ...871...43J,2019ApJ...887..178L}.\par

The first systematic spectral variability analysis of BAL quasar sample was carried out by \citet{1994PASP..106..548B} using multiple epoch observations of 23 BAL quasars. They found BAL trough variations are correlated with the changes in the broadband continuum flux, hence identified the fluctuating ionizing source as the primary cause of BAL variability. 

The studies of \citet{2007ApJ...656...73L} and \citet{2008ApJ...675..985G} found no evidence of variation in continuum flux with the BAL changes and support changing covering fraction of the BAL clouds as a major driver in causing BAL variations. Rigorous analysis of 24 luminous quasars at redshift 1.2 $<$ \zem $<$ 2.9 by \citet{2011MNRAS.413..908C,2012MNRAS.422.3249C,2013MNRAS.429.1872C} over the timescale of 8$-$10 days to 8.7 years\footnote{All the timescales are in rest-frame of quasar} supports both cloud crossing and fluctuating ionizing radiation scenarios based on the variability found only in a portion of BAL troughs and coordinated variability of multiple BAL troughs respectively. \citet{2013ApJ...777..168F} performed the variability analysis in a sample of 291 quasars on a timescale of 1$-$3.7 years and similar to \citet{2010ApJ...713..220G}, they found BAL variation occurs in a portion of BAL troughs with more variation of the narrower regions. They have also reported coordinated BAL trough variabilities in BAL quasars with multiple troughs supporting the changing shielding cloud scenario of BAL variations. Simultaneous multi-epoch spectral and photometric analysis of 22 LoBAL quasars of \aliii and \mgii~absorption lines by \citet{2014MNRAS.440..799V} over a timescale of 10 days to 7.69 years reported no strong correlation between the varying continuum flux and absorption strength. Their analysis favors the line-of-sight cloud crossing scenario as the primary cause of BAL variations. Interestingly, with a sample of 2099 BAL quasars, \citet{2014ApJ...786...42Z} proposed a dust outflow scenario of BAL quasars where they used the slope of near-infrared (NIR) continuum as an indicator for the associated hot dust emissions. They found a moderate correlation between the absorption strength and velocity with slope of NIR continuum. Based on their dust outflow scenario, they suggested dust is intrinsic to the BAL outflows and may contribute to the acceleration provided to BAL clouds \citep[also see][]{1995ApJ...451..510S}. Based on the multi-epoch spectroscopic data set of 6250 quasars from SDSSDR-10, \citet{2015ApJ...814..150W} found a high correlation among the variability of multiple BAL troughs associated with same ions or the same troughs of different ions. They also observed appearance and disappearance of BAL troughs co-occur with the dimming and brightening of the continuum flux, respectively, indicating varying ionizing radiation scenario as a major cause for driving these extreme variations of BAL troughs. \citet{2015MNRAS.454.3962H} presented the variability study of 188 Sloan Digital Sky Survey Data Release 7 (SDSSDR-7) BAL quasar over a timescale ranging from 0.001$-$3 years and found a mild correlation between the variation in continuum luminosity at 1500~\AA\ and variation in the spectral index. They also reported a mild negative correlation between the variation in EW and \NSmgii-based black hole mass and a strong positive correlation between the maximum velocity of outflow and Eddington ratio. Their findings suggest that variabilities are governed by the accretion onto the central engine. \citet{2017MNRAS.469.3163M} explored the emerging and disappearing behavior in 471 SDSS BAL quasars covering a time range of 0.1$-$5.25 years. They reported 14 disappearing BALs with a BAL disappearance rate of 2.3$^{+0.9}_{-0.7}$ within 1.73$-$4.62 years and 18 emerging BALs at a BAL emergence rate of 3.0$^{+1.0}_{-0.8}$ in 1.46$-$3.66 years. Their variability analysis supports both the ionization change and the cloud crossing scenarios. Recent study of \citet{2018A&A...616A.114D} isolates a set of 73 disappearing \civ BAL troughs over the timescale of $\sim$ 3.1 years. Their analysis indicates that the multiple BAL troughs in a BAL quasar vary in a coordinated manner, and the coordination persists over the larger radial distances and hence demands a global mechanism to explain such variations. \citet{2018ApJ...862...22R} presented a detailed analysis of 105 emerging BAL quasars. They found emerging BAL quasars can be represented as BALs of smaller balnicity indices, shallower depths, larger velocities, and smaller widths. They found a rate of coordinated variation of 68.3\% in the multiple BAL troughs at different velocities in the same quasars facing the same fluctuating ionizing radiation. To probe the link between the continuum flux and \civ absorption line strength variation, \citet{2019MNRAS.486.2379V} utilized a multi-epoch spectral data of 78 BAL quasars in the Stripe 82 regions.  The author has discovered a strong correlation between the varying continuum and the absorption trough of shallower depths. On the other hand, no correlation between the continuum and the deeper absorption troughs indicates a significant role of line saturation effects.\par

BAL variability analysis has been performed with a variety of ions, on a range of timescales and different sample sizes. The cause of variability is possibly due to cloud crossing, fluctuating ionizing radiation, and/or a combination of both the scenario. Hitherto, the root cause for the extreme variation i.e., appearance and disappearance of BAL troughs is debated. In this paper, we, for the first time, aim to contrast the properties of appearing and disappearing BAL quasars to explore for the leading cause responsible for such extreme variation of BAL troughs. With this motivation we, (i) provide a new set of appearing (i.e., non-BAL to BAL transition) BAL quasars built up by comparing the data set of SDSSDR-7, SDSSDR-12, and SDSSDR-14, (ii) present a detailed variability study of this new appearing BAL sample, (iii) establish a comprehensive comparison of this appearing BAL quasar sample with an existing large sample of disappearing BAL quasars presented by \citet{2018A&A...616A.114D}.

The current work is organised as follows. In Section 2, we describe how we compiled the sample for our analysis. In Section 3, we outline our data analysis. In Section 4, we present the results. In Section 5, we present the discussions of our findings, followed by the conclusion in Section 6.

\section{The Sample}
\label{sample}
In this section, we describe the selection of appearing and disappearing BAL quasar samples. The  appearing BAL quasar sample is compiled from the comparison of already available quasar catalogs whereas the disappearing BAL quasar sample is directly taken from the  \citet{2018A&A...616A.114D} paper.

\subsection{Initial selection of appearing BAL quasar sample}
\label{prelim_appearingsample}
The initial appearing BAL quasar sample utilized in the present work is constructed based on the quasar properties presented in the quasar catalogs of SDSSDR-7 \citep[][]{2010AJ....139.2360S}, SDSSDR-12 \citep[][]{2017A&A...597A..79P}, and SDSSDR-14 \citep[][]{2018A&A...613A..51P}. The SDSSDR-7 catalog contains 105783 quasars with their detailed properties presented in \citet{2011ApJS..194...45S}. From the analysis of \citet{2011ApJS..194...45S}, 99569 SDSSDR-7 quasars were classified as non-BAL quasars (hereafter NBQ-DR7) with BAL flag 0. The SDSSDR-12 catalog comprises of 297301 quasars, out of which 15177 are BAL quasars (hereafter BQ-DR12) with a non-zero value of Balnicity index (BI) of \civ BAL trough. The remaining 282124 quasars have either BI = 0 or have BAL flag = 0 and constitute a non-BAL DR12 quasar catalog (hereafter NBQ-DR12). On the other hand, SDSSDR-14 quasar catalog has a total of 526357 quasars out of which 21877 quasars have been assigned a non-zero BI value of \civ BAL troughs (BQ-DR14).\par

The process of building the appearing BAL quasar sample can be split into two steps. In step one, we combined the BAL quasars of BQ-DR12 and BQ-DR14 catalogs after discounting the repeated entries in RA and DEC between these two catalogs, which have the same epoch of observation. It is worth noting that 743 BAL quasars in the BQ-DR12 sample for the same epoch of observation are reported as non-BAL quasars in the SDSSDR-14 catalog. Hence, the amalgamation of BQ-DR12 and BQ-DR14 catalogs resulted in 22620 (=21877$+$743) unique BAL quasars and is designated as the BQ-DR12/14 catalog. We then cross-correlated the BQ-DR12/14 BAL quasar catalog with the  NBQ-DR7 catalog using a search radius of 2 arc-seconds. The cross-correlation resulted in a sample of 834 appearing BAL sources uniquely designated as non-BAL quasars in the early epoch SDSSDR-7 catalog and BAL quasars in the later epoch SDSS-12/DR14 catalog (hereafter DR7-DR12/14 appearing BAL sample).\par

In step two, we searched for the sources classified as non-BAL quasars in the SDSSDR-12 catalog but are BAL quasars in the SDSSDR-14 catalog.  For this purpose, we cross-correlated the SDSSDR-12 non-BAL quasar (i.e., NBQ-DR12) catalog with the SDSSDR-14 BAL quasar (i.e., BQ-DR14) catalog within two arc-seconds of search radius. Our search resulted in a sample of 141 sources, which we refer to as the sample of DR12-DR14 appearing BAL quasars. Furthermore, we also noted that among them, seven sources are already members of our DR7-DR12/14 appearing sample of 834 sources (i.e., they remained as a non-BAL quasar in SDSSDR-12 data, and the BAL appeared in the SDSSDR-14 data), while 11 sources were BAL quasar in the SDSSDR-7 data. As a result, DR12-DR14 appearing sample adds new 123 sources (i.e., 141$-$7$-$11), resulting in our total sample of 957 (i.e., 834$+$123) appearing BAL quasars.\par

The spectral coverage of sources in SDSSDR-7 spans a range of 3800$-$9200 \AA\, while the sources in SDSSDR-12/DR14 have spectral coverage of 3600$-$10400 \AA. Therefore, to ensure a proper spectral coverage allowing the detection of \civ BAL troughs for the sources common in the SDSSDR-7 and SDSSDR-12/14 catalogs, we applied a filter based on emission redshift ( \zem), which excludes the sources beyond 1.68 $< $ z$_{em} <$ 4.39 range. This constraint on the z$_{em}$ reduces our sample from 957 to 901 quasars.

Further, to avoid the uncertainty in the analysis due to the inclusion of noisy spectra, we included only those sources with SNR$_{1700} >$ 6 \citep[see][]{2009ApJ...692..758G} in at least two epochs of observation. Out of the 901 sources, we found that 499 sources have SNR$_{1700} >$ 6 spectral data available for at least two epochs. Our candidate sample of 499 quasars is selected based on the comparisons between the catalog data of SDSS DR7, 12, and 14. In order to focus on the quasars exhibiting an appearance of \civ BAL troughs, we require further analysis of the available spectra. The final samples of the study i.e., appearing sample and appearing pristine sample are discussed in Section~\ref{subsec:bal_variability}. Our sample compilation procedure is summarised in Table~\ref{tab:table_summary_selection}.\par

\subsection{The comparison sample of disappearing BAL quasars}
\label{subsec:dis_sample}
To draw a comparison between appearing BAL  quasars and  disappearing BAL quasars, we chose the largest sample of 67 disappearing BAL quasars  presented in \citet{2018A&A...616A.114D} for which spectral data is readily available from SDSS-I/II/III. Since this sample is derived from SDSS-I/II/III programs similar to our study, any bias arising in the comparison due to the intrinsic spectral properties (i.e., spectral coverage and resolution) can be neglected. The classification of their 67 disappearing BAL quasars was performed based on the two-sample $\chi^{2}$ test on the flux value of the two available epochs of each BAL quasars. To minimize the random occurrence of the disappearance of BAL troughs, the authors required the probability of the test (P$_{\chi}^{2}$) to be P$_{\chi}^{2} \leq$10$^{-4}$. Their analysis isolated 73 such BAL troughs in the spectra of 67 BAL quasars. They labeled it as \pfour sample. Among this \NSpfour\ sample, 52 BAL troughs had P$_{\chi}^{2} \leq$10$^{-8}$, and the authors labeled this sample as \NSpeight. However, upon visual inspection, only 30 BAL troughs in \peight had no residual absorption in their later epoch observations, and this sample was tagged as \NSdispri\ sample.\par
     
\begin{table}
  \renewcommand{\thetable}{\arabic{table}a}
  \begin{adjustwidth}{-0.5cm}{}
  \scriptsize
   \caption{Summary of sample selection that resulted  in the compilation of appearing BAL sample.}
    \centering
    \begin{tabular}{clr}
    \hline
    (1)& SDSSDR-7 quasars with BAL flag$^{a}$ $=$ 0 (NBQ-DR7)                   & 99569  \\
    (2)& SDSSDR-7 quasars with BAL flag$^{a} \neq$ 0                            & 6214   \\    
    (3)& SDSSDR-12 quasars with BAL flag$^{b}$ = 0                              & 282124 \\
       & or BI$=$0 (NBQ-DR12)                                                   &        \\
    (4)& SDSSDR-12 quasars with BI $>^{b}$ 0 (BQ-DR12)                          & 15177  \\
    (5)& SDSSDR-14 quasars with BI$^{c}$ $>$ 0 (BQ-DR14)                        & 21877 \\
    \hline
    Step-1 &                                                                    &\\
    (6)& (4) and (5) excluding repeated entries (BQ-DR12/14)                    & 22620\\
    (7)& (1) and (6) within 2 arc-seconds (DR7-DR12/14 appearing)                & 834\\
    \hline
    Step-2 &                                                                    &\\
    (8)& (3) and (5) within 2 arc-seconds (DR12-DR14 appearing)                  & 141 \\
    (9)& (7) and (8) within 2 arc-seconds                                       & 7 \\
    (10)& (2) and (8) within 2 arc-seconds                                      & 11 \\
    (11)& (7) $+$ (8) $-$ (9) $-$ (10)                                          & 957 \\
    \hline
    (12)& (11)  and 1.68 $< $z$_{em} <$ 4.39                                & 901 \\
    (13)& (12) and multi-epoch SN$_{1700}^{\mathrm{spectra}} >$ 6 spectra   & 499 \\
    \hline
    \multicolumn{3}{l}{$^{a}$ \citet{2011ApJS..194...45S}, $^{b}$ \citet{2017A&A...597A..79P}; $^{c}$\citet{2018A&A...613A..51P};}\\
    \end{tabular}
    \label{tab:table_summary_selection}
  \end{adjustwidth}
\end{table}

\begin{table}
\addtocounter{table}{-1}
\renewcommand{\thetable}{\arabic{table}b}
\scriptsize
\begin{adjustwidth}{-0.5cm}{}
\caption{Summary of various subsamples involved in the present study.}
\begin{tabular}{lccccl}
\toprule 
 Sample                &     Number of        & Number of      &  Number of  & Criteria\\
 Type                  &     BAL quasars      & MJD pairs      &  complexes  &        \\
\midrule                                                                    
\maincomplex           &         195          &  198           &    255      &  A\\
\emii                  &         107          &  107           &    116      &  A$+$B$+$C\\
\emiipri               &         59           &  59            &    62       &  D$+$E$+$F\\
\dismain               &         51           &  51            &    53       &  G$+$B$+$C$+$H
\\
\hline
\multicolumn{5}{l}{A: BI$=$0 in Epoch-1 from present analysis;  ~B: \fracew > 5$\times$ \NSfracewsig;}\\ 
\multicolumn{5}{l}{C: P$_{\chi}^{2} \leq$10$^{-4}$; D: Subsample of \emii BAL sample; E: P$_{\chi}^{2} \leq$10$^{-8}$;}\\
\multicolumn{5}{l}{F: Visual inspection; G: BI$=$0 in Epoch-2 from present analysis;}\\
\multicolumn{5}{l}{H: Common in \pfour of \citet{2018A&A...616A.114D}.}\\
\end{tabular}
\label{tab:sample_summary}
 \end{adjustwidth}
\end{table}


\section{Data Analysis}
\label{sec:analysis}

\subsection{Continuum fitting}
\label{subsec:conti}
We used multi-epoch spectra that are normalized by an estimated continuum model to study the variations in BAL characteristics. To determine the proper continuum levels, we first corrected the main-sample spectra for Galactic extinction using a Milky Way extinction model \citep[see][]{1999PASP..111...63F} for R$_{V} =$ 3.1. The A$_{V}$ values were taken from the \citet[][]{1998ApJ...500..525S}.  The night-sky lines from the data were removed using the ``BRIGHTSKY'' mask flag provided by SDSS.  We then translated observed wavelengths to the rest-frame using redshifts from \citet[][]{2010MNRAS.405.2302H}. We constructed the underlying continuum by fitting a power-law model that is intrinsically reddened using the SMC-like reddening model from \citet[][]{2003ApJ...594..279G} to only relatively line-free (RLF) windows of the spectral regions: 1250-1350~\AA, 1700-1800~\AA, 1950-2200~\AA, 2650-2910~\AA, 3950-4050~\AA.  The RLF windows were selected to be relatively free from emission and absorption lines considering the composite quasar spectra of \citet[][]{2001AJ....122..549V}. The three power-law continuum-model parameters are thus the power-law normalization, the power-law spectral index, and the intrinsic absorption coefficient.  To exclude the data points that deviate from the fit by more than 3$ \sigma$, we apply an iterative sigma-clipping algorithm using a non-linear least-squares fit.\par

\begin{figure*}
\includegraphics[scale=0.56,trim=90 0 0 0, clip]{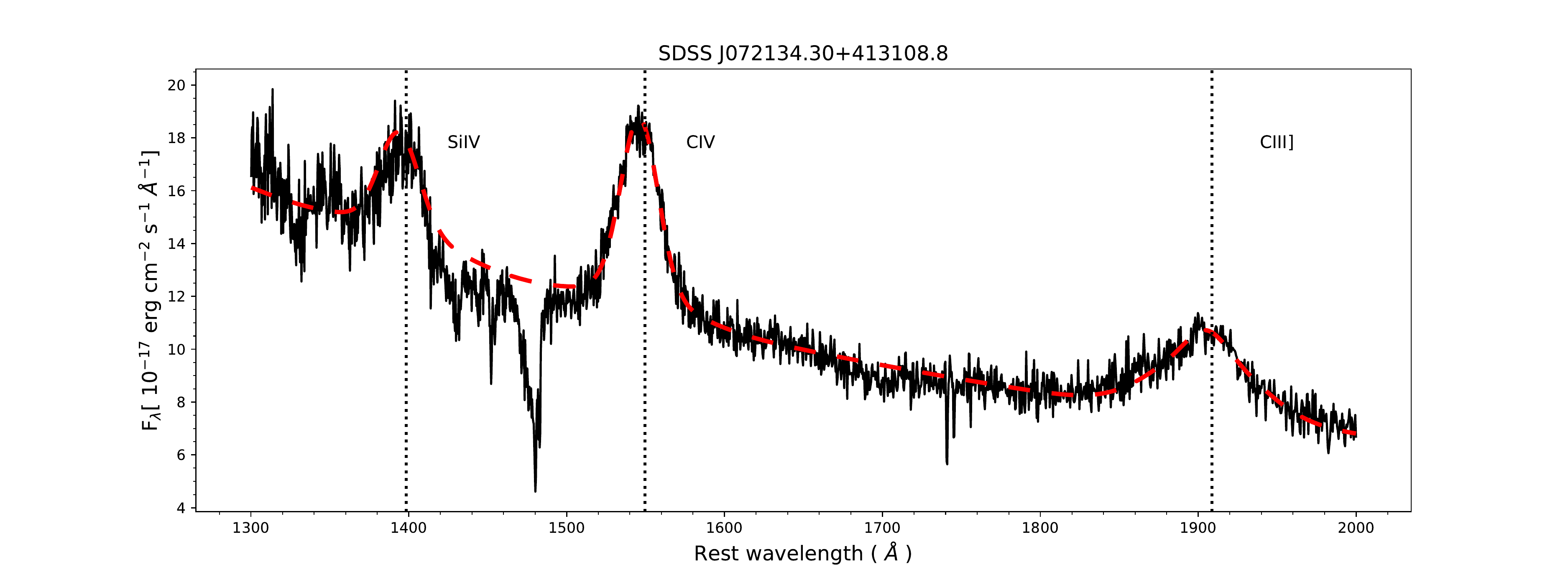}
\caption{The continuum fit (red dashed curve) for the spectrum of SDSS-J072134.30$+$413108.8 modeled by a combination of a reddened power law, double Gaussian for \civ and \siiv emission lines and a single Gaussian for \ciii emission line. Three vertical lines (black dotted) correspond to the quasar's \sivab, \civab, and \ciiia emission lines.}
\label{fig:conti1}
\end{figure*}

Besides, we employed a double Gaussian function for the fitting of \civ and \siiv emission line doublets and a single Gaussian function for \ciii emission line. Here we have tied the width and redshift of the Gaussian components of these species.  The continuum model generated after the simultaneous fitting of the reddened power-law, two double Gaussians for \civ and \NSsiiv, and a single Gaussian \ciii for emission line is shown in Fig.~\ref{fig:conti1} for the spectrum of SDSS-J072134.30$+$413108.8. To evaluate the uncertainties over the continuum fit, we performed the Monte Carlo simulations by randomizing the flux in each pixel with a random Gaussian deviate associated with the uncertainty in the pixel for 100 realizations. We fit the continuum to these 100 realizations and adopted their standard deviation as the uncertainty of the continuum fit which also accounts for the uncertainty of the emission lines.\par

\subsection{BAL identification}
\label{subsec:BAL_iden}
To formally quantify a BAL trough, \citet{1991ApJ...373...23W} defined the Balnicity Index (BI), which measures the strength of all the \civ absorption troughs in a spectrum over a velocity range of 3000 to 25000\kms blue-ward of the quasar \civ emission. According to this BI definition, the normalized flux of each absorption trough needs to be below 0.9 for at least 2000 \NSkms.\par
In this work, we have used spectral regions between 3000 to 30000 \kms blue-ward of the quasar \civ emission to exclude the regions which are prone to be affected by the variability caused in quasars \civ and \siiv emission lines \citep[see][]{2012ApJ...757..114F}. To identify the multiple BAL troughs in each quasar spectrum, we have used a modified form of BI for each \civ BAL trough and defined as:

\begin{equation}
  \textrm{BI}_{\rm trough} \equiv \int^{\rm v_{min}}_{\rm v_{max}} \bigg[1-\frac{f(\rm v)}{0.9}\bigg] C \ \rm dv
  \label{eq.BI}
\end{equation} 

where $f(\rm v)$ is the normalized flux density at velocity v, and $C$ is a constant, which is 1 when the normalized flux is below 0.9 for at least 2000~\kms and 0 otherwise. The upper (v$_{max}$) and lower (v$_{min}$) limit on velocities are defined as the maximum and minimum velocity of the trough within which the normalized continuum level is continuously below 0.9 for at least 2000 \NSkms. An absorption trough with BI$_{\rm trough} >$ 0 is considered as BAL absorption trough.\par
 
Since the appearance of the BAL troughs is evaluated in the multi-epoch spectra of the BAL quasars, v$_{max}$ and v$_{min}$ of each \civ BAL trough is derived using all the available spectra of each BAL quasar. A BAL trough may split into two small absorption troughs, or small troughs may merge to form a broad trough from one epoch to another epoch of observation. Therefore, to quantify the variability of absorption troughs between two epochs of BAL quasar spectra, we considered absorption complexes as presented in \citet{2013ApJ...777..168F} and \citet{2018ApJ...862...22R}.
\begin{figure}
\hspace{-0.3in}
\includegraphics[scale=0.55]{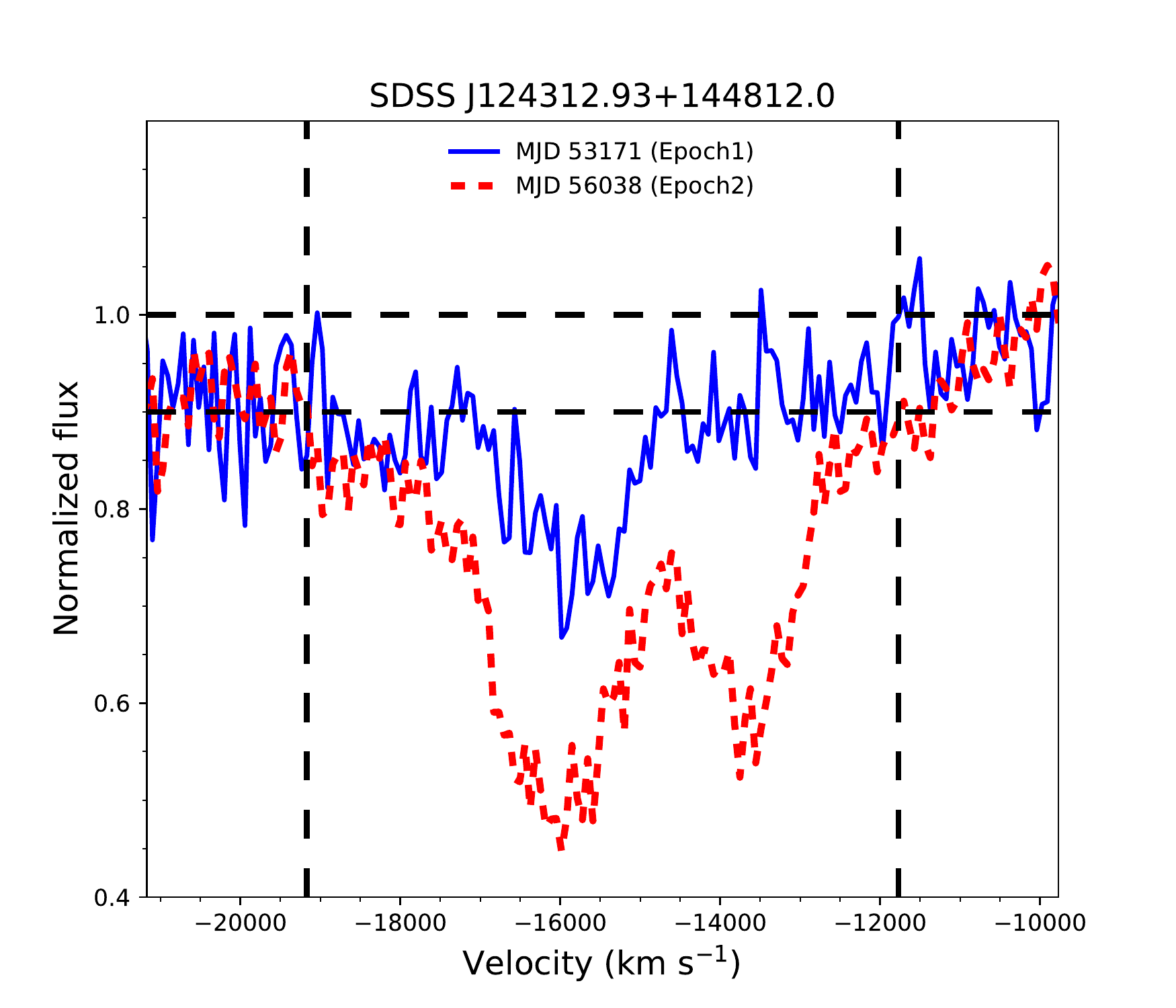}
\caption{Demonstration of determining the lower (v$_{min}$) and upper (v$_{max}$) limits of an absorption complex using the multi-epoch spectra of SDSS J124312.93$+$144812.0. Two vertical dashed lines indicate the lower and upper limit of velocity region used for the BAL trough, while two horizontal dashed lines represent the normalized continuum level at 1.0 and 0.9.}
\label{fig:abs_complex}
\end{figure}
If v$_{min}$ and/or v$_{max}$ of a BAL trough, detected in one epoch of a BAL quasar intercepts inside the BAL trough found in other epoch of the same BAL quasar, the minimum of the lower velocities of the BAL troughs from both the epochs is set as the v$_{min}$ of the absorption complex. Similarly, the maximum of the higher velocities of the BAL troughs is taken as the v$_{max}$ of the absorption complex. However, in the case of no BAL trough detection in one of the epochs, the lower and higher velocities of the BAL trough detected in the other epoch is taken as the v$_{min}$ and v$_{max}$ of absorption complexes respectively. In this way, each absorption complex contains all the BAL troughs, which may have split into multiple small BAL troughs from one epoch to another epoch of observation for a BAL quasar. In Fig.~\ref{fig:abs_complex}, we show an example of absorption complex found in SDSS J124312.93$+$144812.0, where the BAL trough has increased in strength along with an emergence of a new component in the later epoch as compared to the earlier epoch. However, the variability measurement of this BAL trough has been performed between the v$_{min}$ and v$_{max}$ estimated from the later epoch.
Among the 499 sources (Section~\ref{sample}), 419 sources have 2 epochs, 62 sources have 3 epochs, 13 sources have 4 epochs, 3 source has 5 epochs, 1 source has 6 epochs while 1 source has 7 epochs of spectral data (i.e., a total of 2$\times$419 $+$ 3$\times$62 $+$ 4$\times$13 $+$ 5$\times$3 $+$ 6$\times$1 $+$ 7$\times$1= 1104 spectra). 
As \citet{2018A&A...616A.114D} has only considered the BAL disappearance phenomena happening between two consecutive epochs of observations, we also have imposed the same criterion  in our analysis to ensure a uniform comparison between BAL appearance and disappearance samples.
Therefore, for the sources with spectral data available for more than two epochs, we have treated each consecutive unique pair of MJD as a single source. For instance, J002028.34$-$002914.9 has three epochs of spectra available at MJDs 51816, 51900 and 56979; therefore the combinations of MJDs 51816-51900 and 51900-56979 are considered as two different measurements in our analysis equivalent to two sources. Hence our sample of 499 sources having  1104 spectra resulted in 605 MJD pairs (i.e., 419 $+$ 2 $\times$62 $+$ 3$\times$13 $+$ 4$\times$3 $+$ 5$\times$1 $+$ 6$\times$1 ). 
Among these 605 MJD pairs of 499 BAL quasars, we report 586 BAL complexes among 488 MJD-pairs of 414 BAL quasars. However, in 117 MJD-pairs of 85 BAL quasars, no BAL detection was found in both the epochs of observation and hence excluded them from further analysis. We note that from this 586 BAL complexes, in vast majority of complexes, i.e., 361 complexes in 323 MJD pairs of 270 BAL quasars, we identified BAL troughs with BI $\neq$ 0 in both earlier and later epochs of the observations whereas the study of \citet{2011ApJS..194...45S} characterises them as non-BAL quasars from their earlier epoch observation (even to few visually obvious cases e.g., SDSS J141723.73$+$285522.6). These noted discrepancies between the catalogs and the present work may be emerging from the difference in the methods employed for fitting the continuum. Besides, the catalog presented by \citet{2011ApJS..194...45S} is not explicitly tuned for BAL detection. Since the present work focuses on the appearing behavior of \civ BAL complexes, we have excluded these BAL complexes in further analysis. The remaining 225 complexes in 198 MJD pairs of 195 BAL quasars have BI $\neq$ 0 only in the later epochs.\par

We  also note that among these 195 BAL quasars, 51 BAL quasars have an additional non-appearing BAL component together with appearing BAL troughs. Although our initial selection criteria based on BI are tuned to identify only BAL quasars with one or more appearing BAL troughs, the difference in the BAL complex identification procedure results in BAL quasars with multiple BAL components where one of the component is  non-appearing. We have validated these multiple components through visual inspection. As none of the previous studies on BAL variability, including that of \citet{2018A&A...616A.114D}, have  excluded these additional components from their analysis, we also retain these 51 sources in the present study to allow for a uniform comparison between the appearing and disappearing BAL quasars. However, the reader is reminded about the caveat that our catalog based selection procedure preferentially selects BAL quasars where all the BAL components appear in the later epoch. We refer the sample of this 255 BAL complexes in 198 MJD pairs of 195 BAL quasars as {\it ``Preliminary''} sample. In Table~\ref{tab:643_all_BAL} we have listed the physical properties of this \maincomplex sample, where the values of bolometric luminosity (log $L_{\rm{bol}}$), black hole mass (log $M_{\odot}$) and Eddington ratio (log $\frac{L_{\rm{bol}}}{L_{\rm{Edd}}}$) are based on the study of \citet{2011ApJS..194...45S}.\par

\begin{table*}
 \begin{minipage}[10]{140mm}
\caption{Physical parameters of \maincomplex sample of 195 BAL quasars with 198 MJD pairs.}
\begin{tabular}{ccccccccccc}
\toprule 
BAL quasar          & MJD-pair & z$_{em}$  & $M_i$     & log $L_{\rm{bol}}$& log $M_{\rm BH}$ & log $\frac{L_{\rm{bol}}}{L_{\rm{Edd}}}$ & $\Delta$t\\
                    &          &          & (mag)     &  (erg s$^{-1}$)  &  (\(M_\odot\))  &                                    & (years)  \\
\midrule
J000330.19$+$000813.2 & 52519-55478 & 2.59  & -27.61  & 46.99$\pm$0.01 & 9.42$\pm$0.47  & -0.53  & 2.26 \\   
J001641.17$+$010045.2 & 52518-55480 & 3.05  & -27.68  & 46.89$\pm$0.02 & 9.58$\pm$0.10  & -0.79  & 2.01 \\   
J002028.34$-$002914.9 & 51900-56979 & 1.94  & -26.68  & 46.76$\pm$0.01 & 9.48$\pm$0.05  & -0.82  & 4.74 \\   
J002845.77$+$010648.3 & 52930-55181 & 2.47  & -26.63  & 46.65$\pm$0.02 & 9.79$\pm$0.30  & -1.26  & 1.78 \\   
J005157.24$+$000354.7 & 51913-52201 & 1.95  & -27.87  & 47.27$\pm$0.01 & 9.73$\pm$0.07  & -0.56  & 0.27 \\
---                   & ---         & ---  & ---    & ---            &---            & ---          & --- \\
\hline
\multicolumn{8}{l}{{ Note.} The entire table is available  in on-line version. Only a portion of this table is shown here,}\\
\multicolumn{8}{l}{to display its form and content.}\\
\end{tabular}
\label{tab:643_all_BAL}
\end{minipage}
\end{table*}


\subsection{BAL trough properties measurement}
\label{subsec:bal_properties}

For quantifying the properties of BAL troughs associated with the absorption complexes in each BAL quasar spectrum, we measured the EW (in \AA), centroid velocity (in \NSkms), width (in \NSkms), and depth below the normalized continuum.\par

To measure the EW in \AA\ and its uncertainty from the normalized spectra, we used equations 1 and 2 of \citet{2002ApJ...574..643K}:
\begin{equation}
  \textrm{EW} = \sum_i \left( 1 - \frac{F_i}{F_c} \right) B_i,
  \label{eq:EW}
\end{equation}
and,
\begin{equation}
  \sigma_{\textrm{EW}} = \sqrt{\left[\frac{\Delta F_c}{F_c} \sum_i\left(\frac{B_iF_i}{F_c}\right) \right]^2 + \sum_i\left(\frac{B_i\Delta F_i}{F_c}\right)^2}.
  \label{eq:sigEW}
\end{equation}
where $F_i$ is the normalized flux with associated error $\Delta F_i$ at $i$th pixel and $F_c$ is the mean of the continuum normalized flux and $\Delta F_c$ is the corresponding error on mean. $B_i$ is the width of a wavelength bin measured in units of angstrom. The summation is carried out over the wavelength range set by v$_{min}$ and v$_{max}$ as explained in Section~\ref{subsec:BAL_iden}.\par

The centroid velocity, v$_{cent}$, of a BAL trough is defined as the mean of the normalized flux weighted velocities in the window within v$_{min}$ and v$_{max}$. The width of a BAL trough, $\Delta$v, is taken as the difference between the v$_{max}$ and the v$_{min}$. The normalized depth, d$_{\scaleto{\rm BAL \rm}{3.5pt}}$ of each BAL trough is computed by taking the mean of the normalized fluxes inside the window over which BAL features are identified.\par

\subsection{Identification of appearing/disappearing BAL troughs}
\label{subsec:bal_variability}
To identify BAL appearance/disappearance from one epoch to another, we computed the change in the EW and the corresponding uncertainty as:

\begin{equation}
  \Delta \textrm{EW} = \textrm{EW}_2 - \textrm{EW}_1 \qquad \sigma_{\Delta \textrm{EW}} = \sqrt{\sigma^2_{\textrm{EW}_2} + \sigma^2_{\textrm{EW}_1}}.
  \label{eq:deltaEW}
\end{equation}

where the subscripts `1' and `2' correspond to the earlier and later epoch of observation, respectively. Similarly, we computed the fractional EW variations and the corresponding uncertainties using the following equations:

\begin{equation}
\frac{\Delta \rm EW}{\langle \rm EW \rangle}=\frac{\rm EW_2-\rm EW_1}{(\rm EW_2 + \rm EW_1)\times 0.5}
    \label{eq.d_EW1}
\end{equation}

\begin{equation}
\sigma_{\frac{\Delta \rm EW}{\langle \rm EW \rangle}}=\frac{4\sqrt{\rm EW_{\rm 2}^2\rm \sigma^2_{\textrm{EW}_1}+\rm EW_{\rm 1}^2\rm \sigma^2_{\textrm{EW}_2}}}{(\rm EW_{\rm 2} + \rm EW_{\rm 1})^2}
    \label{eq.d_EW2}
\end{equation}

where $\langle \rm EW \rangle$ is the average of the EWs measured between the two epochs; $\rm \sigma_{\textrm{EW}}$ is the associated uncertainty in the EW values.  The fractional change in EW ( \fracew) depicts the significance of the change in the absorption compared to the size of changing feature.\par

Similar to the change in EW, the change in the normalized depths with associated uncertainties of the absorption complexes are also computed as:

\begin{eqnarray}
    \Delta d_{\scaleto{\rm BAL \rm}{3.5pt}} = d_{\scaleto{\rm BAL,2 \rm}{4pt}} - d_{\scaleto{\rm BAL,1 \rm}{4pt}} \\
    \sigma_{\Delta d_{\scaleto{\rm BAL \rm}{3.5pt}}} = \sqrt{\sigma^2_{d_{\scaleto{\rm BAL,2 \rm}{4.5pt}}} + \sigma^2_{d_{\scaleto{\rm BAL,1 \rm}{4.5pt}}}}. 
  \label{eq:deltaDBAL}
\end{eqnarray}

To probe the role of ionizing source, we evaluated fractional flux variation of the line free continuum at 1700 \AA~(~\fraccon), and corresponding uncertainty using the following equations:
\begin{equation}
\frac{\Delta \rm F_{\rm cont,2}}{\langle \rm F_{cont,1} \rangle}=\frac{\rm F_{\rm cont,2}-\rm F_{\rm cont,1}}{(\rm F_{\rm cont,2} + \rm F_{\rm cont,1})\times 0.5}
    \label{eq.F1}
\end{equation}

\begin{equation}
\sigma_{\frac{\Delta \rm F_{\rm cont}}{\langle \rm F_{cont} \rangle}}=\frac{4\sqrt{\rm F_{\rm cont,2}^2\rm \sigma^2_{\textrm{F}_{cont,1}}+\rm F_{\rm cont,1}^2\rm \sigma^2_{\textrm{F}_{cont,2}}}}{(\rm F_{\rm cont,2} + \rm F_{\rm cont,1})^2}
    \label{eq.F2}
\end{equation}

where $\rm F_{\rm cont1}$ and $\rm F_{\rm cont2}$ represent the power-law continuum flux at 1700 \AA~ for the two epochs respectively; $\rm \sigma_{\textrm{F}_{cont}}$ is the associated flux uncertainty.\par

In order to ensure a proper non-BAL to BAL transition of our \maincomplex sample (e.g., see Section~\ref{subsec:BAL_iden}), we have applied stringent criteria on the significance of appearance. In addition to appearance of BAL trough at 5$\sigma$~ (i.e., \fracew > 5$\times$ \NSfracewsig), following \citet{2018A&A...616A.114D}, we have used two sample $\chi^{2}$ test on the flux values corresponding to a particular velocity window over which BAL appearance is detected in a particular MJD-pair. We have required the probability P$_{\chi^{2}}$ associated with the test to be $<$ 10$^{-4}$ to avoid the random occurrence of the transition. Among the set of 225 BAL complexes from \maincomplex sample, 116 BAL complexes in 107 MJD pairs of 107 BAL quasars have shown the appearance of BAL troughs at 5$\sigma$~ with P$_{\chi^{2}} <$ 10$^{-4}$. We refer this sample as {\it ``Appearing''} sample. Among these 107 BAL quasars, the present study isolates 84 new appearing BAL quasars, while 23 have already been reported in \citet{2018ApJ...862...22R}.

It is important to note that, even though our \emii sample consists of BAL complexes with absorption changing at more than 5$\sigma$ with P$_{\chi^{2}} <$ 10$^{-4}$, residual absorption that does not amount to a non-zero BI$_{trough}$ may exist in a few cases of the earlier epoch spectra. Therefore, to isolate the definite cases of appearing complexes with negligible residual absorption in the earlier epochs, we have put even more stringent requirement i.e., P$_{\chi^{2}} <$ 10$^{-8}$ in addition to a visual inspection of the BAL complexes from the \emii sample. Based on this inspection from among the \emii BAL sample and having P$_{\chi^{2}} <$ 10$^{-8}$, we have identified a pristine sample of 62 BAL complexes in 59 MJD pairs of 59 BAL quasars (hereafter {\it ``Appearing pristine''}).\par

As discussed in Section~\ref{subsec:dis_sample}, to perform a comparison of our appearing BAL quasars with the disappearing BAL quasars, we made use of the largest sample of 67 disappearing BAL quasars from \citet{2018A&A...616A.114D}. Using the information provided in the Table A.1 of \citet{2018A&A...616A.114D}, we assembled the multi-epoch spectra of their \pfour sample and subjected them to the same analysis as that of the current work (as of Section~\ref{sec:analysis}). In total, we identified 96 disappearing \civ BAL complexes in 65 MJD pairs of 65 BAL quasars. Out of these 96 BAL complexes, our analysis yielded 63 \civ BAL troughs in 54 BAL quasars, which had undergone a disappearance of their BAL troughs at 5$\sigma$ and have P$_{\chi^{2}} <$ 10$^{-4}$. It is important to note that among these 63 disappearing BAL troughs, only 53 BAL troughs in 51 BAL quasars are common with \pfour sample of 73 BAL troughs in 67 BAL quasars \citep[see Table A.2 of][]{2018A&A...616A.114D}. Unlike \citet{2018A&A...616A.114D}, we only included region between 3000$-$30000\kms blue-ward of quasar's \civ emission line. In addition, we examined the disappearance of BAL absorption complexes as compared to the analysis of the disappearance of different BAL components considered in \citet{2018A&A...616A.114D}. As a result, few disappearing BAL troughs were included as single BAL troughs in our analysis  as against multiple troughs in the analysis of \citet{2018A&A...616A.114D}. Besides, different continuum normalization procedures  in these two studies may contribute to the observed difference in BAL trough detection. Therefore, the seen discrepancy of 20 BAL troughs between \citet{2018A&A...616A.114D} and this  study can be attributed to the difference in the BAL trough analysis in these studies. However, to avoid any disagreement based on the difference in the analysis employed in the two studies, we only considered those 53 \civ BAL troughs in 51 BAL quasars which are common in \pfour sample of \citet{2018A&A...616A.114D}. We refer to this sample as {\it ``Disappearing''} sample. A summary of the various samples is presented in Table~\ref{tab:sample_summary}.\par

\begin{table*}
\scriptsize{
\caption{Details of the \NSemii, \NSemiipri, and \dismain samples of \civ BAL absorption complexes.}
\begin{tabular}{lllrlllrrccccc}
  \toprule
  BAL quasar & MJD pair  & \multicolumn{1}{c}{\delt} & \multicolumn{1}{c}{\fraccon}  & \multicolumn{1}{c}{$\Delta$v} & \multicolumn{1}{c}{v$_{min}$} & \multicolumn{1}{c}{v$_{max}$} & \multicolumn{1}{c}{\delew} & \multicolumn{1}{c}{\fracew} & \multicolumn{1}{c}{\deld} & \multicolumn{1}{c}{\avgd} & Type \\
             &           & year         &         & \multicolumn{1}{c}{\kms}      & \multicolumn{1}{c}{\kms}     & \multicolumn{1}{c}{\kms}     &  \multicolumn{1}{c}{\AA}      &         &        &   &  & \\
  \midrule
SDSS J000330.19$+$000813.2 & 52519$-$55478 & 2.26 &  $-$1.31$\pm$0.75 & 2689.41  & 13573.67   & 16263.08  & 1.54   $\pm$0.23  &  0.79   $\pm$0.15  &  0.14 $\pm$0.02   &   0.16$\pm$0.01   &  \emiipri \\    
SDSS J000330.19$+$000813.2 & 52519$-$55478 & 2.26 &  $-$1.31$\pm$0.75 & 3016.18  & 19573.16   & 22589.34  & 3.62   $\pm$0.24  &  1.14   $\pm$0.11  &  0.25 $\pm$0.01   &   0.22$\pm$0.01   &  \emiipri \\    
SDSS J002028.34$-$002914.9 & 51900$-$56979 & 4.73 &  $-$0.72$\pm$0.82 & 2462.49  & 17116.74   & 19579.23  & 2.18   $\pm$0.20  &  1.00   $\pm$0.11  &  0.16 $\pm$0.01   &   0.16$\pm$0.01   &  \emiipri \\    
SDSS J002845.77$+$010648.3 & 52930$-$55181 & 1.78 &  $-$0.88$\pm$0.88 & 13291.93 & 12870.93   & 26162.86  & 12.30  $\pm$0.86  &  0.83   $\pm$0.07  &  0.20 $\pm$0.01   &   0.22$\pm$0.01   &  \emiipri \\    
SDSS J004022.40$+$005939.6 & 52261$-$55182 & 2.24 &  0.60   $\pm$0.66 & 6453.83  & 3081.89    & 9535.71   & $-$9.89$\pm$0.64  &  $-$1.31$\pm$0.06  &  $-$0.25$\pm$0.02 &   0.23$\pm$0.01   &  \dismain \\ 
\multicolumn{1}{c}{---}                   & \multicolumn{1}{c}{---}         & \multicolumn{1}{c}{---}  &  \multicolumn{1}{c}{---}             &  \multicolumn{1}{c}{---}          &   \multicolumn{1}{c}{---}   & \multicolumn{1}{c}{---}         &  \multicolumn{1}{c}{---}        & \multicolumn{1}{c}{---}           & \multicolumn{1}{c}{---} & \multicolumn{1}{c}{---} & \multicolumn{1}{c}{---}  \\
\hline
\multicolumn{12}{l}{{ Note.} The entire table is available  in the online version. Only a portion of this table is shown here, to display its form and content.}\\
\label{Table:510_signi_vari_BAL}
\end{tabular}
}
\end{table*}

\section{Results}
\label{Section:result}
In this part, we present comparisons of rest-frame timescales  (Section~\ref{subsec_var_time}), quasar intrinsic parameters (Section~\ref{subsec_var_intrinsic}), BAL trough parameters (Section~\ref{subsec:var_bal_pro}), parameters associated with the continuum (Section~\ref{subsec:continuum_para}), and reddening parameter (Section~\ref{subsec:spectral_reddening}) between the \NSemii, \NSemiipri, and \dismain BAL quasars.\par

\begin{figure}
\hspace{-0.2in}
\includegraphics[scale=0.55]{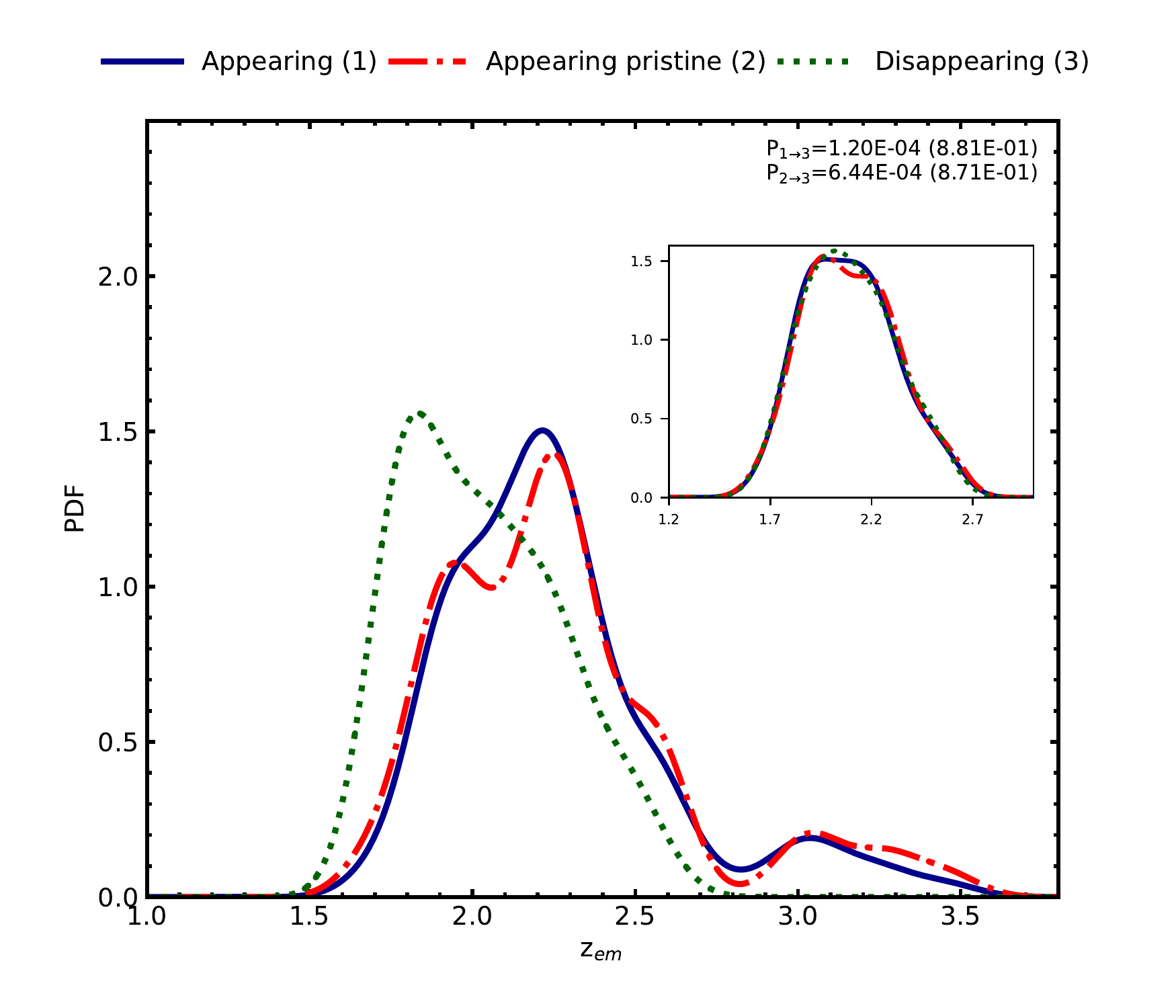}
\caption{KDE distribution of emission redshift generated using Gaussian kernel of fixed bandwidth for \emii (\emph{solid blue}), \emiipri (\emph{dashed red}) and \dismain (\emph{dotted green}) samples. The inset shows the similar plots for the control sample of \emii, \emiipri and \dismain BAL quasars. The p-values computed using the K-S test between \emii and \dismain (P$_{1 \rightarrow 3}$), and \emiipri and \dismain (P$_{2 \rightarrow 3}$) are given in top right corner with corresponding p-values for the control samples in brackets.}
\label{fig:zem_cs}
\end{figure}

In Fig.~\ref{fig:zem_cs}, we present the distributions of \zem\ for \NSemii, \NSemiipri, and \dismain samples. We used a non-parametric kernel density estimation \citep[KDE, ][]{1986desd.book.....S} with a Gaussian kernel of fixed bandwidth to derive the probability density function (PDF) for each distribution mentioned above. We have used the maximum likelihood cross-validation approach to assess an optimal bandwidth value for Gaussian kernels. We performed a two-sample Kolmogorov-Smirnov (K-S) test in these distributions, and the corresponding p-values of the K-S test are given in the top right corner of Fig.~\ref{fig:zem_cs}; where the p-value of the K-S test gives the null probability for the two distributions being identical. 

It can be seen from Fig.~\ref{fig:zem_cs} that the \zem~ distributions for \NSemii\ and \NSemiipri\ BAL quasars differ significantly from the \zem~ distribution of \dismain BAL quasars. The  \NSemii\ and \NSemiipri\ BAL quasars are found to be at higher redshift as compared to the \dismain BAL quasars. The p-values computed using the K-S test between \emii and \dismain (P$_{1 \rightarrow 3}$), and \emiipri and \dismain (P$_{2 \rightarrow 3}$) are given in the top right corner of Fig.~\ref{fig:zem_cs} and both the p-values point to a statistically significant difference between these distributions. 
To minimize the effect of the difference in the redshift distributions in our analysis, we constructed control samples of \emii and \NSemiipri\ BAL quasars by only selecting the \emii and \NSemiipri\ BAL quasars which have a close-by redshift to that of the \dismain BAL quasars. In this process, we perform a one-to-one matching of redshift between the \NSemii, \NSemiipri, and \dismain samples with a tolerance of $\Delta$z$_{em} = \pm$ 0.05. 
In the inset of Fig.~\ref{fig:zem_cs}, we have shown the \zem~ distributions for the control samples of \NSemii, \NSemiipri, and \dismain BAL quasars. Since we are performing one-to-one redshift matching, to avoid selective inclusion of any particular source, we have 1000 times randomized the sequence of objects in \NSemii, \NSemiipri, and \dismain BAL quasar samples keeping their sample size intact. We then performed one-to-one redshift matching between these 1000 randomized samples so that we have 1000 control samples matching in redshift each for \NSemii, \NSemiipri, and \dismain BAL quasar samples. With these 1000 control samples, we then calculated P$_{1 \rightarrow 3}$ and P$_{2 \rightarrow 3}$ values and adopted their median as the final values of P$_{1 \rightarrow 3}$ and P$_{2 \rightarrow 3}$ for the control samples and are given in parentheses of Fig.~\ref{fig:zem_cs}. As can be inferred from the p-values given in the parentheses of Fig.~\ref{fig:zem_cs}, the redshift distributions of our control samples are similar. In the following sections, we present the comparisons of \NSemii, \NSemiipri, and \dismain BAL quasars and verify the significance of our comparisons using the control samples.\par

\subsection{Comparison of rest-frame timescales}
\label{subsec_var_time}
\begin{figure}
\hspace{-0.2in}
\includegraphics[scale=0.55]{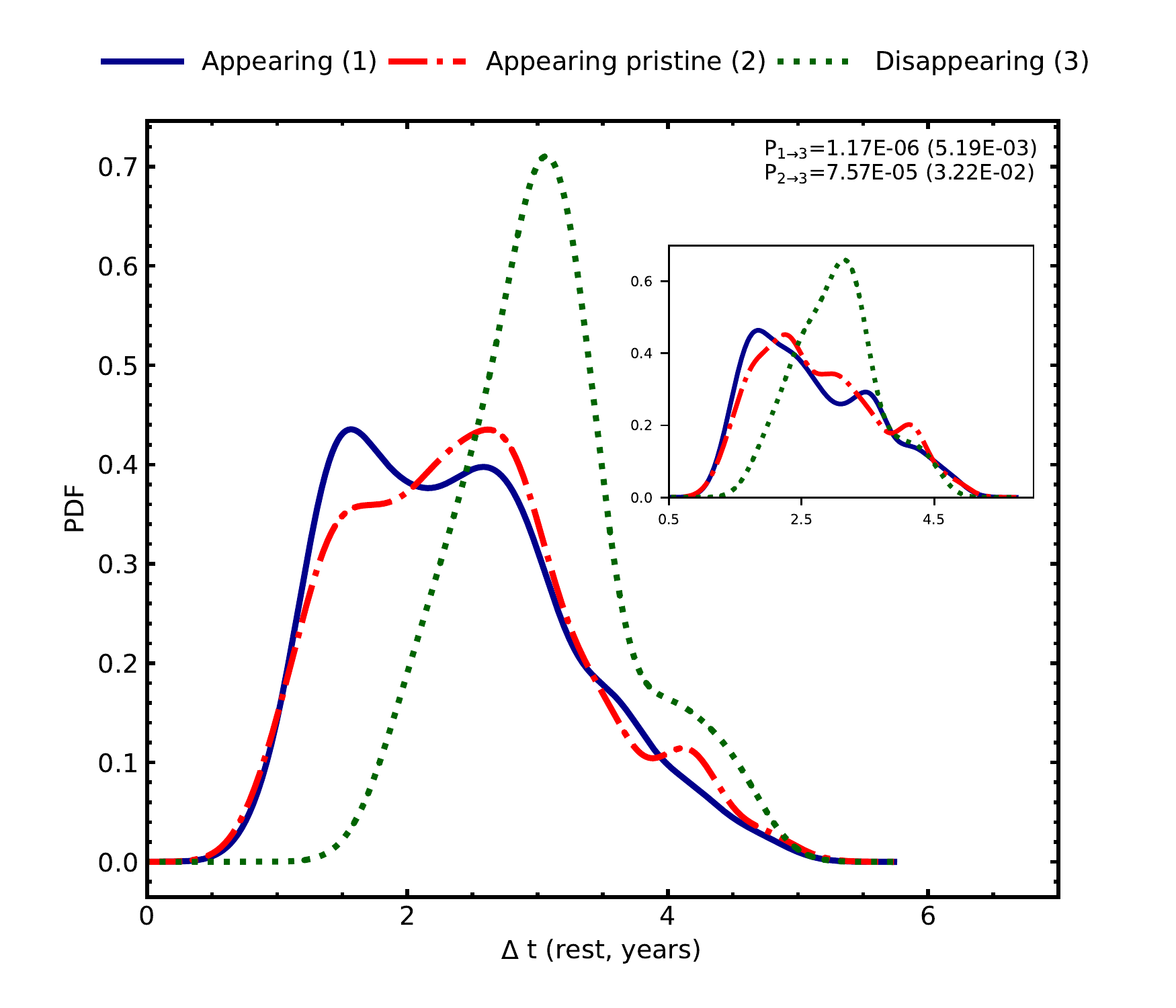}
\caption{Distribution of rest-frame timescale between the two epochs derived with KDE using Gaussian kernel of fixed bandwidth for \emii (\emph{solid blue}), \emiipri (\emph{dashed red}) and \dismain (\emph{dotted green})  samples. In the inset the similar distributions for the control samples of \emii, \emiipri and \dismain BAL quasars are given. The p-values computed using the K-S test between \emii and \dismain (P$_{1 \rightarrow 3}$), and \emiipri and \dismain (P$_{2 \rightarrow 3}$) are given in top right corner with their corresponding values for control samples in brackets.}
\label{fig:time_zem_comp}
\end{figure}

In this section, we compare the rest-frame timescales for the BAL appearance and disappearance. Fig.~\ref{fig:time_zem_comp} presents the distribution of the rest-frame timescale  for the \NSemii, \emiipri and \dismain BAL quasar samples. It is evident from Fig.~\ref{fig:time_zem_comp} that the \emii and \emiipri BAL transitions significantly occur on smaller timescales compared to the \dismain BAL transitions with p-values of the K-S test being less than 5\%. Given that there is a significant difference in the redshift distributions between these samples, it is unsurprising to see the difference in the rest-frame timescales. However, the distribution of rest-frame timescales for the control samples of \emii and \emiipri BAL quasars also considerably differ from the control sample of \dismain BAL quasars (note the P-values in the parentheses of Fig.~\ref{fig:time_zem_comp}). They follow a similar trend as that of the main samples, with the \emii BAL quasars being found at smaller timescales than the \dismain BAL quasars. The difference in the timescales for both main and control samples of \NSemii, \NSemiipri, and \dismain BAL quasars point to a dissimilar temporal sampling of spectra in these samples. Hence from Fig.~\ref{fig:time_zem_comp}, we infer that the probed timescales of \emii BAL quasars are smaller compared to the probed timescales of \dismain BAL quasars.

\subsection{Comparison of quasar intrinsic properties}
\label{subsec_var_intrinsic}

\begin{figure}
\hspace{-0.5in}
\includegraphics[scale=0.55,trim=15 0 0 0,clip]{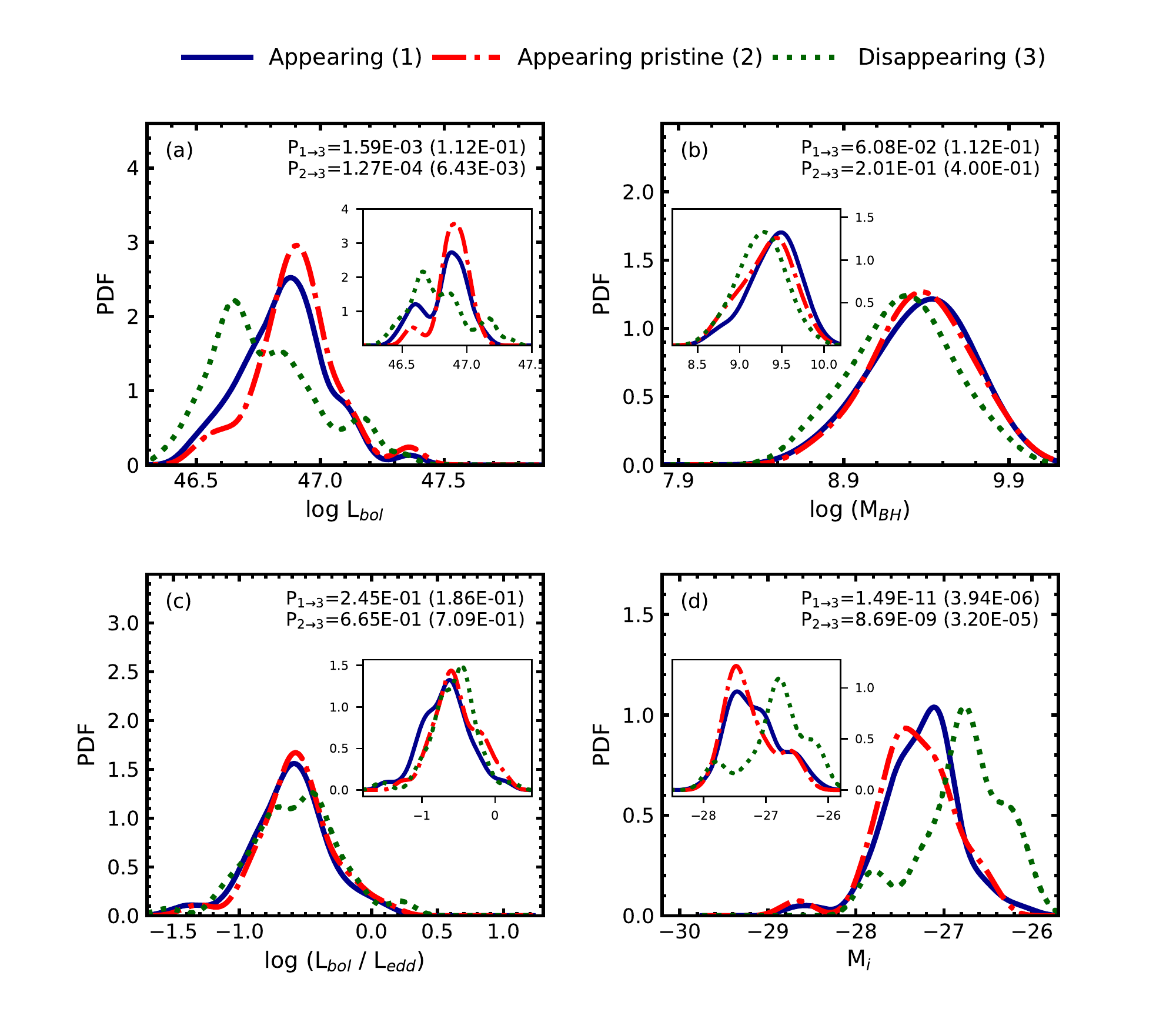}
\caption{KDE distribution of (a) bolometric luminosity, (b) black hole mass, (c) Eddington ratio and (d) {\it i}-band absolute magnitude for \emii (\emph{solid blue}), \emiipri (\emph{dashed red}) and \dismain (\emph{dotted green}) samples generated using Gaussian kernel of fixed bandwidth are drawn. The insets in each panel gives the similar distributions for the control sample of \emii, \emiipri and \dismain. The estimated p-values of the K-S test between \emii and \dismain (P$_{1 \rightarrow 3}$), and \emiipri and \dismain (P$_{2 \rightarrow 3}$) with their corresponding values for the control samples in brackets are shown in the top right corner of each panel.}
\label{fig:intrinsic_parameter}
\end{figure}

It is known that accretion disk winds are very much sensitive to the Eddington ratio, defined as the ratio of bolometric luminosity to the Eddington luminosity, which further depends on the black hole (BH) mass \citep{2004ApJ...616..688P}. Therefore, the BAL quasar properties are expected to depend on these parameters \citep[see][]{2002ApJ...569..641L,2007ApJ...665..990G}. To probe whether the nature of the ionizing sources in the appearing and disappearing BAL quasars are similar, we in Fig.~\ref{fig:intrinsic_parameter} compare the quasar parameters such as (a) bolometric luminosity, (b) BH mass, (c) Eddington ratio, and (d) i-band absolute magnitude for \NSemii, \NSemiipri, and \dismain BAL quasars. We adopt these quasar properties from the study of \citet{2011ApJS..194...45S}. We wish to emphasize that as the quasar intrinsic parameters are adopted from the \citet{2011ApJS..194...45S} catalog, the intrinsic parameters are reflective of the physical conditions in the first epoch of spectroscopic observation.\par

From Fig.~\ref{fig:intrinsic_parameter}~(a and d), we note that the distributions of bolometric luminosity and absolute i-band magnitude for \emii and \emiipri BAL quasars differ significantly from the \dismain BAL quasars with p-values of K-S test for the \emii and \dismain (P$_{1 \rightarrow 3}$) and \emiipri and \dismain (P$_{2 \rightarrow 3}$) BAL samples consistent with less than 5\%. It is to be noted that a similar trend is found for the control samples of \emii and \dismain BAL quasars (mainly between \emiipri and \dismain BAL quasars), which are at similar redshift as can be seen from the KDE distributions in the inset of Fig.~\ref{fig:intrinsic_parameter}~(a and d). This implies that the \emii BAL quasars are more luminous and brighter compared to the \dismain BAL quasars. However, Fig.~\ref{fig:intrinsic_parameter}~(b and c) shows no difference in the BH mass and Eddington ratio distributions of \NSemii, \NSemiipri, and \dismain BAL quasars (similarly for the control samples, especially between \emiipri and \dismain BAL quasars), suggesting that central environments of \emii and \dismain BAL quasars are alike.\par

It is worth noting that the Eddington ratio's calculations involve the measurements of black hole masses, which further requires the estimations of full width at half maximum (FWHM) of quasar emission lines. Since our sources are at high redshift (with median z$_{em} \sim$ 2), the BH mass estimates are predominately based on \civ emission line \citep[see][]{2011ApJS..194...45S}. The \civ emission lines exhibit remarkable displacements towards the blue side \citep{1982ApJ...263...79G}, possibly due to strong outflows; hence the measurements of FWHM and virialized velocity dispersion are inaccurate in such cases. As a result, \civ based BH mass estimations are systematically biased compared to the BH mass derived from other low-ionization quasar emission lines (such as H$\beta$ and \mgii). Therefore, any conclusion inferred based on the Eddington ratio and BH mass must be handled with caution. To be conservative, we will not interpret the results based on the BH mass and Eddington ratio in this study.\par

\subsection{Comparison of BAL-profile Properties}
\label{subsec:var_bal_pro}
\begin{figure*}
\includegraphics[scale=0.65]{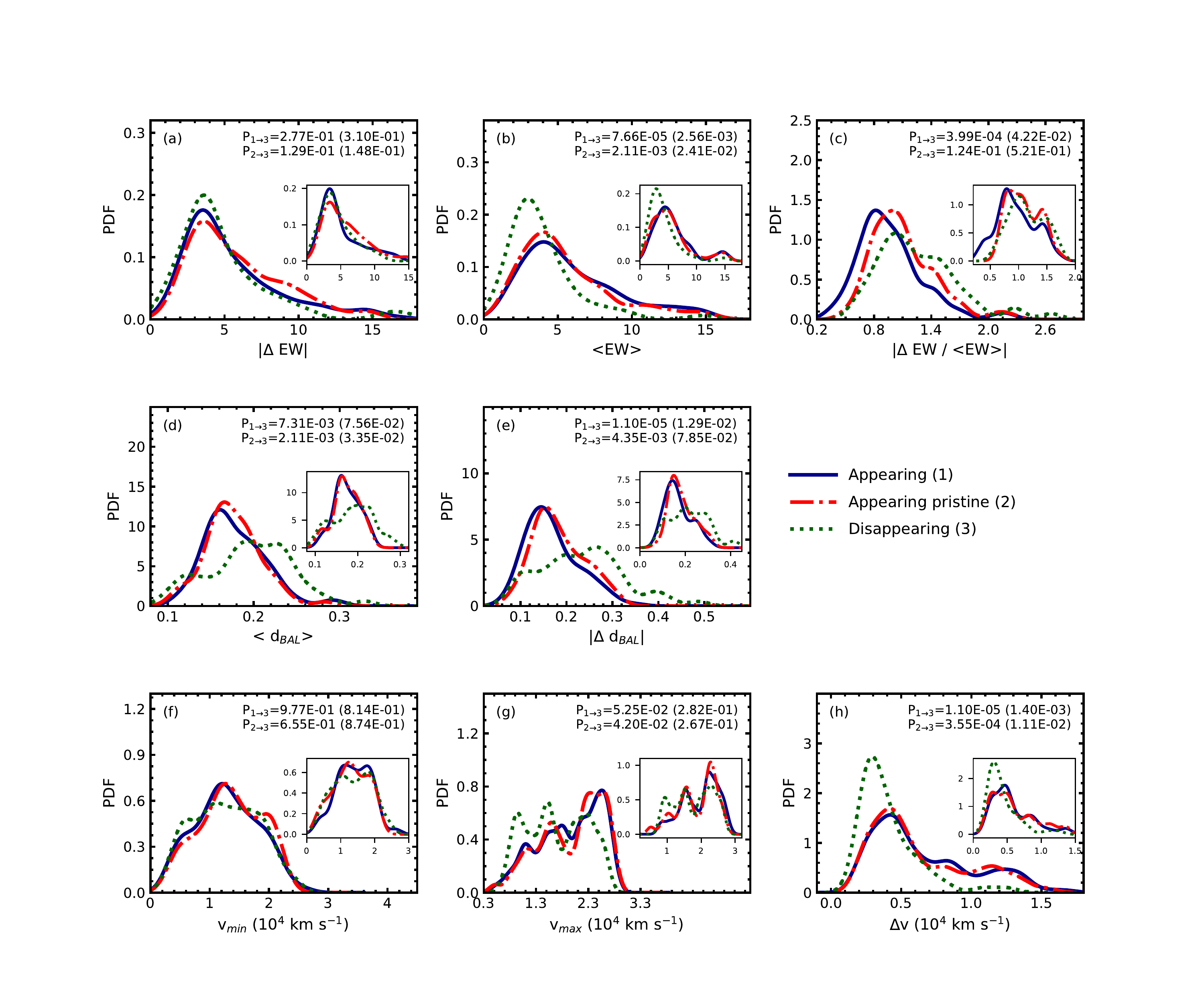}
\caption{KDE distribution of (a) absolute change in EW, (b) average EW, (c) absolute fractional change in EW, (d) average normalized BAL depth, (e) absolute change in normalized BAL depth, (f) minimum outflow velocity, (g) maximum outflow velocity, and (h) width of BAL trough for \emii (\emph{solid blue}), \emiipri (\emph{dashed red}) and \dismain (\emph{dotted green}) BAL complexes derived with Gaussian kernel of similar bandwidth. In the inset of each panel similar distributions for the control sample of \emii, \emiipri and \dismain BAL quasars are drawn. The p-values derived from the K-S test between \emii and \dismain (P$_{1 \rightarrow 3}$), and \emiipri and \dismain (P$_{2 \rightarrow 3}$) with their corresponding values for the control samples in brackets are mentioned at the top right corner of each panel.}
\label{fig:absorption_comp}
\end{figure*}

In this section, we compare the BAL trough properties such as the average EW (\NSavgew), average normalized trough depth (\NSavgd), change in normalized trough depth (\NSdeld), maximum/minimum velocity of outflow, and trough width (\NSdelv) among the \NSemii, \NSemiipri, \NSdismain, and their control samples. In Fig.~\ref{fig:absorption_comp}, we show the KDE distributions of these parameters. 
The p-values computed using the K-S test for the two pairs of distributions i.e. \emii and \dismain (P$_{1 \rightarrow 3}$,) and \emiipri and \dismain (P$_{2 \rightarrow 3}$) are also shown in each panel while the  values in the parentheses are the p-values evaluated for the corresponding control samples. 

From Fig.~\ref{fig:absorption_comp}~(a), it can be discerned that the distributions of ~|\NSdelew|  are similar for \NSemii, \emiipri and \dismain BAL troughs in the main as well as in the control samples as can also be concluded from the p-values. However, the distributions of \NSavgew\ (Fig.~\ref{fig:absorption_comp}~b) and  |\NSfracew| (Fig.~\ref{fig:absorption_comp}~c)   show significant difference between the main samples of \NSemii, \NSemiipri, and \dismain BAL quasars with p-values less than 5\%.  It is evident from Fig.~\ref{fig:absorption_comp}~(b) that the \emii BAL quasars have larger \avgew compared to the \dismain BAL quasars. As applied to the control sample, the difference in \NSavgew\ distributions are still significant whereas the |\NSfracew| distributions do not show a marked difference. Also, the distribution of |\NSfracew| is similar for \emiipri and \dismain BAL quasars (with P$_{2 \rightarrow 3} <$ 5\%). Since by definition, \emiipri BAL sample is the conservatively reliable sub-sample of \emii BAL sample with systematically larger values of |\NSfracew| of BAL complexes, we conclude from Fig.~\ref{fig:absorption_comp}~(c) that the absolute fractional variation of EW is similar in appearing and disappearing BAL quasars.

It is evident from Fig.~\ref{fig:absorption_comp}~(d, e, f, g and h) that the distributions of \NSavgd, |\NSdeld|, v$_{max}$, and \delv for the main samples of \emii and \emiipri differ significantly from the \dismain sample (see P$_{1 \rightarrow 3}$ and P$_{2 \rightarrow 3}$) whereas there is no significant difference in the distributions of v$_{min}$. However,  for the control samples of \NSemii, \NSemiipri, and \dismain BAL quasars, the difference in the distribution of v$_{max}$ is not significant (see p-values in brackets of Fig.~\ref{fig:absorption_comp}~g). From Fig.~\ref{fig:absorption_comp}~(d, e, and h) it can be inferred that the \emii BAL quasars have shallower and wider BAL troughs as compared to the \dismain BAL quasars. Same can be verified from their corresponding control samples mainly between \emiipri and \dismain BAL quasars. 

It is interesting to note the opposing trends in the distributions of BAL depths and BAL widths. Since BAL equivalent width is the function of BAL depth  and BAL width, the similarity in the ~|\NSdelew| distribution between the \emii  and \dismain BAL quasars can be understood as a combination of the opposing trends in the |\NSdeld| and \delv distributions. Correspondingly, the opposing trends in the \avgd and \delv distributions should also result in a similar distribution of \NSavgew\ among \emii  and \dismain BAL quasars. However, the  \NSavgew\ distributions differ significantly between the two samples. The trend in \NSavgew\ distributions closely resemble the trend in the \delv distributions. The apparent discrepancy in the distributions of \NSavgd, \NSdelv, and  \NSavgew\ may be explained by the lower values of \avgd\ as compared to |\NSdeld|. In the case of extremely varying BAL troughs like appearing/disappearing troughs, it is expected that the  \NSavgd\ values are significantly lower than the |\NSdeld| values (in the ideal BAL disappearing/appearing case, |\NSdeld| = 2$\times$\NSavgd). Fig.~\ref{fig:absorption_comp}~(d and e) also point to lower values of \avgd as compared to |\NSdeld|. We surmise that this relatively smaller contribution of \NSavgd\ to the \NSavgew\, as opposed to a larger contribution of |\NSdeld| to |\NSdelew|, may explain why the opposing trends in \avgd and \delv distributions do not combine to result in a similar \NSavgew\ distribution for the \emii and \dismain BAL quasars (Fig.~\ref{fig:absorption_comp}~b).\par 
Further, using Spearman correlation test, we searched for correlations between \fracew and other BAL-trough parameters (i.e \NSavgew, \NSavgd, \NSdeld, v$_{min}$, v$_{max}$, and \NSdelv) for the \NSemii, \emiipri and \dismain samples of BAL quasars \citep{2007ApJ...656...73L,2008ApJ...675..985G,2011MNRAS.413..908C,2013ApJ...777..168F,2014MNRAS.440..799V,2018A&A...616A.114D}. Our analysis shows no significant correlation between \fracew and  BAL-trough parameters, \NSavgew, \NSavgd, v$_{min}$, v$_{max}$, and \delv for \NSemii, \NSemiipri, and \dismain samples. Although previous BAL variability studies have found significant correlations between  \fracew and  BAL-trough parameters, the lack of significant correlation between these parameters in this study may be because this study focuses on the extreme cases of BAL variability, which only spans a narrow range of \NSfracew. However, we note that there is a very strong positive and highly significant ($>$ 99.9\%) correlation between \fracew and \deld for all the three BAL samples.\par

\begin{figure*}
\includegraphics[scale=0.65]{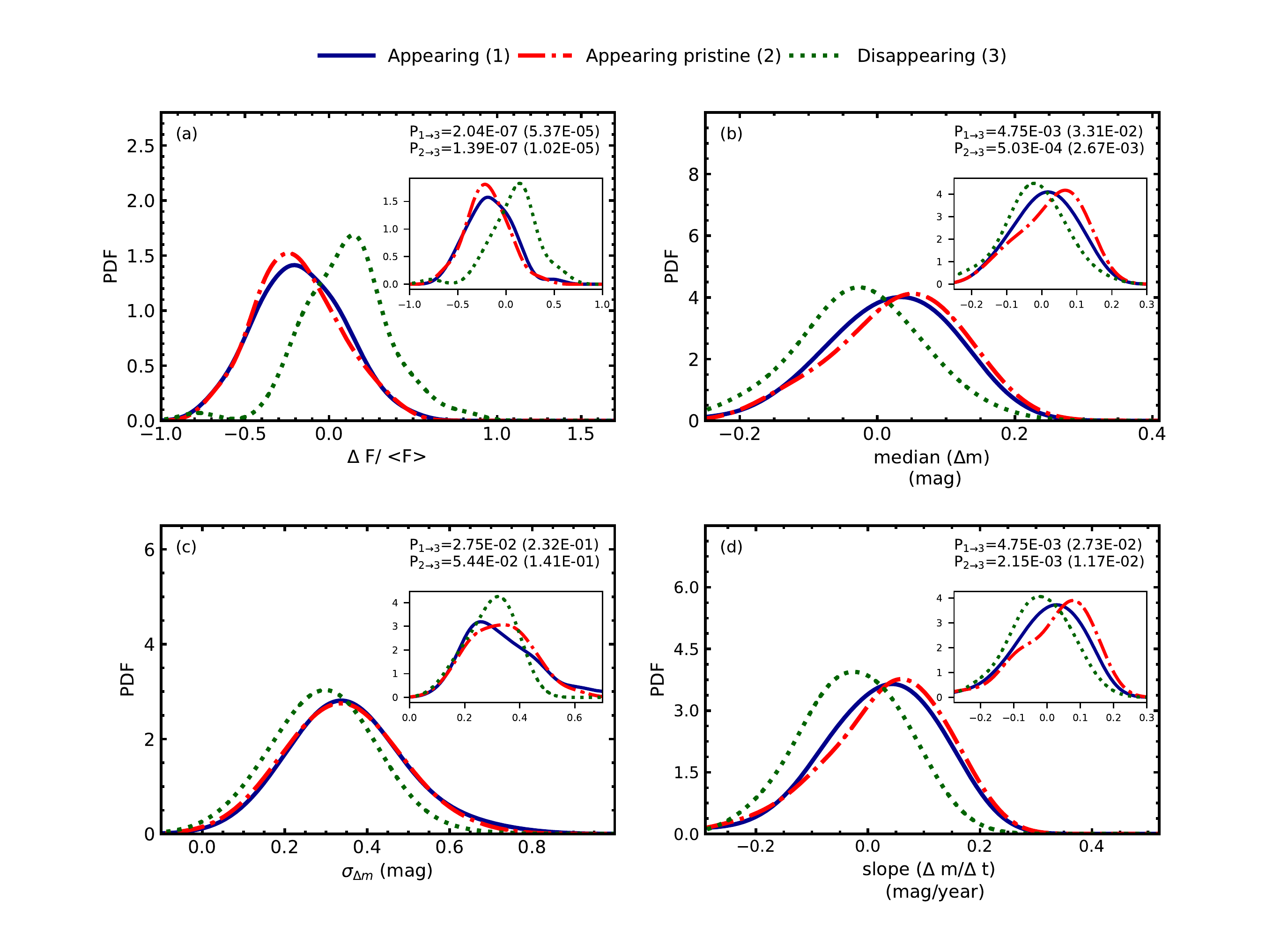}
\caption{Distribution of (a) change in continuum flux at 1700~\AA\, (b) median of change in V-band magnitudes, (c) standard deviation of change in V-band magnitudes and (d) slope of change in V-band magnitude with time for \emii (\emph{solid blue}), \emiipri (\emph{dashed red}) and \dismain (\emph{dotted green}) samples constructed with KDE using Gaussian kernels of same bandwidth. The inset in each panel shows the similar distributions for the control sample of \emii, \emiipri and \dismain BAL quasars. The p-values of the K-S test applied on \emii and \dismain (P$_{1 \rightarrow 3}$), and \emiipri and \dismain (P$_{2 \rightarrow 3}$) with their corresponding values for the control samples in brackets are labeled at the top right corner of each panel.}
\label{fig:continuum_comp}
\end{figure*}
\subsection{Comparison of continuum properties}
\label{subsec:continuum_para}
As discussed in Section~\ref{sec:intro}, change in the photo-ionizing conditions inside the central engine can cause the variation in the optical depths of ions in the BAL clouds and therefore, can contribute to the large change in the EW of BAL troughs. Previous studies based on the coordinated variabilities of BAL troughs of multiple ions at the same velocity \citep{2013ApJ...777..168F,2015ApJ...814..150W,2017MNRAS.469.3163M,2018MNRAS.473L.106L} and of the same ion at different velocities \citep{2012MNRAS.422.3249C,2018A&A...616A.114D,2018ApJ...862...22R} confirmed that similar mechanism is accountable for such coordinated variations of the BAL troughs. The transverse motions of the BAL clouds across the line of sight are less likely to explain such variation unless the BAL outflows are confined to smaller distances relative to the ionizing source. In addition, a significant correlation between the change in continuum and change in absorption strength of the BAL troughs \citep{2015MNRAS.454.3962H,2018MNRAS.473L.106L,2019ApJ...883...30L,2019MNRAS.486.2379V} supports the ionization driven scenario of BAL trough variations. \citet{2015MNRAS.454.3962H} found that in 60.5 percent of their BAL troughs, the decrease in EW is accompanied by the brightening of the continuum, and spectra turned bluer during this phase. The study of \citet{2015ApJ...814..150W} showed that the appearance and disappearance of BAL troughs are accompanied by the dimming and brightening of the continuum. The authors suggested that in such an extreme scenario where the variability amplitude of the BAL trough is very high, the ionic column density is highly sensitive to ionizing radiation. Similarly, \citet{2017ApJS..229...22H} found that in 80 percent of BAL QSOs, the trough variabilities are governed by the change in the ionizing continuum. However, the trends of BAL variabilities, whether it decreases or increases in response to the change in the continuum, depends on the ionization parameter of the absorbers. The study of \citet{2019MNRAS.486.2379V} revealed that the correlation between the continuum radiation and fractional EW variation strengthen when line saturation effects in BAL troughs are eliminated.\par

To probe whether the appearance or disappearance of BAL troughs is caused by the fluctuations produced in the continuum emission coming from the central environment, we used the variation of the continuum flux at 1700 \AA\ (\NSfraccon) as a proxy for the ionizing continuum fluctuations from one epoch to another. Also, to further confirm the effect of fluctuating ionizing continuum on the appearance and disappearance of BAL troughs, we searched for the Johnson's V magnitude light curves from the Catalina Real-Time Transient Survey \citep[CRTS;][]{2009ApJ...696..870D} for our appearing and disappearing samples \citep[see][]{2014MNRAS.440..799V,2014MNRAS.440.2474W,2016MNRAS.455..136V}. The light curves obtained for our sources were taken between April 2005 to December 2013, and each object necessarily has at least one spectroscopic epoch intercepting the photometric time-span over which the light curve is obtained. To quantify the continuum variation from the CRTS light curves, we adopted the procedure used in the study of \citet{2014MNRAS.440..799V}. On a given night, CRTS provides four observing frames 10 min. apart for each object. Since we are dealing with long term continuum variations, similar to \citet{2014MNRAS.440..799V}, we averaged the V-band magnitudes of these four frames to eliminate the large dispersion in the light curves. We measured the V-band magnitude difference (\NSdelm) among all possible photometric MJD pairs (\NSdelt) available in each source's light curve. We fitted an error-weighted least-square straight line in \NSdelm$-$\delt space and obtained the slope (\NSslopedelm). To mark the direction and amplitude of continuum variation, we used the median and standard deviation (\NSsigdelm) of \delm values, and slope of \NSdelm$-$\delt plane (\NSslopedelm) from all light curves.\par

Fig.~\ref{fig:continuum_comp} gives the distribution of \NSfraccon, \NSmeddelm, \sigdelm and \slopedelm for \NSemii, \emiipri and \dismain samples. The p-values of the K-S test applied between \emii and \dismain (P$_{1\rightarrow 3}$), \emiipri and \dismain (P$_{2 \rightarrow 3}$) samples are marked on the top right of each panel. The distributions of \NSfraccon, \NSmeddelm, and \slopedelm for the \emii and \dismain samples are clearly separated and are distributed on either side of zero (Fig.~\ref{fig:continuum_comp}~a, b, and d). However, the distributions of \NSsigdelm, which signifies the amount of V-band magnitude variation, in Fig.~\ref{fig:continuum_comp}~(c) for \NSemii,  \NSemiipri, and \dismain samples show no considerable difference implying that the amplitude of V-band magnitude variations for the BAL appearance and disappearance are similar. The KDE distributions of \fraccon for \emii and \emiipri samples significantly differ from the \dismain sample (Fig.~\ref{fig:continuum_comp}~a). From these distributions, it is evident that the appearance of BAL troughs is more likely to be accompanied by a decrease in continuum flux, whereas the disappearance of BAL troughs is accompanied by an increase in continuum flux. This is in agreement with the study of \citep{2015ApJ...814..150W}. Also, the significant difference  in Fig.~\ref{fig:continuum_comp}~(b and d) for \meddelm and \slopedelm distributions of \NSemii, \emiipri and \dismain samples further supports the anti-correlation between continuum changes and BAL trough changes. We note the similar contrasting trends of these continuum properties for our control samples of \NSemii, \NSemiipri, and \dismain BAL quasars (see the inset KDE distributions of Fig.~\ref{fig:continuum_comp}~a, b, and d).\par

The PDF curves in Fig.~\ref{fig:continuum_comp}~(b and d) show that the increase in V-band magnitude facilitates the appearance of BAL troughs while a decrease in V-band magnitude supports the disappearance of BAL troughs. It is interesting to note that the distributions of \fraccon are markedly different and well separated among the \NSemii, \NSemiipri, and \dismain samples compared to the distributions of \meddelm and \NSslopedelm. The \fraccon were estimated from the same spectra that were used to measure the \fracew and hence span the same time interval for the continuum and BAL trough measurements. On the other hand, the values of \meddelm and \slopedelm are measured using CRTS light curves that do not span the same time interval of observations over which transitions in the BAL troughs have occurred. Additionally, as magnitudes are defined in the logarithmic unit of flux values, \meddelm and \slopedelm distributions for the \emii and \dismain samples are not as detached as the \fraccon distribution.\par

To investigate in what manner and amount do the fluctuation in the ionizing parameters correlates with the fractional variation of EW, we searched for the correlations using the Spearman correlation test between the \fracew and these continuum parameters. We did not find any significant correlation between \fraccon and \fracew for either of our subsamples. The absence of correlation between the \fracew and the fluctuating continuum is possibly expected since we are only dealing with the extreme cases, unlike the previous studies.\par

\begin{figure}
\includegraphics[scale=0.5]{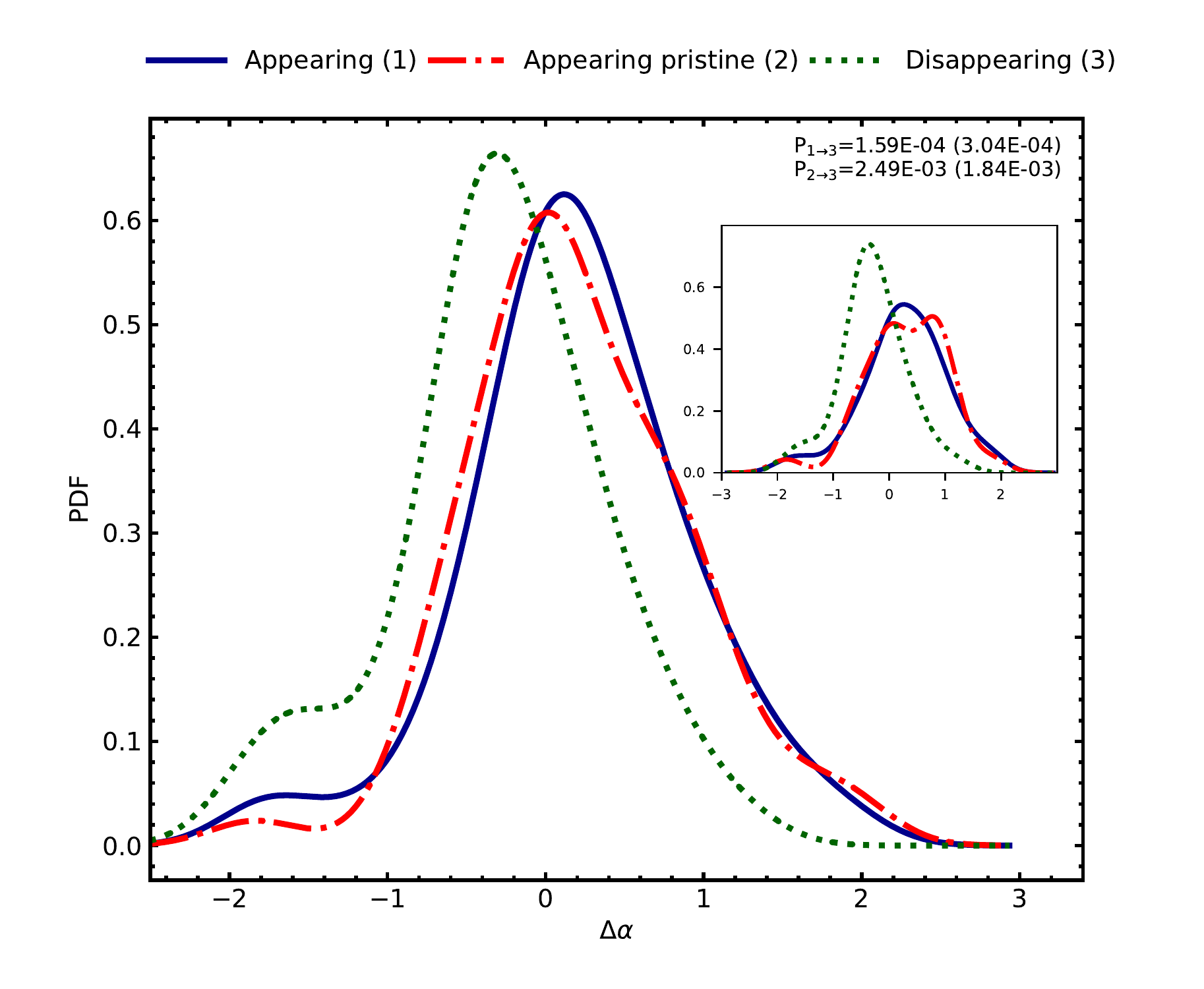}
\caption{Distribution of change in spectral index generated with KDE taking Gaussian kernels of similar bandwidth for \emii (\emph{solid blue}), \emiipri (\emph{dashed red}) and \dismain (\emph{dotted green}) samples. The inset shows the similar distribution for the control sample of  \emii, \emiipri and \dismain BAL quasars. The p-values of the K-S test estimated for \emii and \dismain (P$_{1 \rightarrow 3}$), and \emiipri and \dismain (P$_{2 \rightarrow 3}$) with corresponding values for the control samples in brackets are given at the top right corner.}
\label{fig:alpha_comp}
\end{figure}

We also investigated the nature of variation of spectral index estimated from the reddened power law fitting (See Section~\ref{subsec:conti}) on each spectrum of appearing and disappearing BAL quasars. In Fig.~\ref{fig:alpha_comp}, we present the distribution of change in spectral index (\NSdelalp) from one epoch to another for \NSemii, \emiipri and \dismain samples. Interestingly, we found \delalp distributions of \emii and \emiipri samples are remarkably different from \dismain BAL quasar sample (note P$_{1 \rightarrow 3}$ and P$_{2 \rightarrow 3}$ of Fig.~\ref{fig:alpha_comp}). Similar contrasting difference in the \delalp distributions among the control samples of \NSemii, \NSemiipri, and \dismain BAL quasars can also be inferred from the KDE distributions  in the inset panel of Fig.~\ref{fig:alpha_comp}.\par
It is clear from Fig.~\ref{fig:alpha_comp} that the \delalp values for the \emii and \emiipri BAL quasars are distributed above zero (i.e., \delalp $>$ 0 and spectral softening), while for the \dismain BAL quasars, they are distributed below zero (i.e., \delalp $<$ 0 and spectral hardening). This result, together with the previous observation that BAL appearance/disappearance is accompanied by continuum dimming/brightening, implies that the quasars considered in this study follow the "bluer when brighter" trend for continuum variations. The bluer-when-brighter trend for continuum in quasars has been extensively explored in the past studies  \citep{2012ApJ...744..147S,2014ApJ...792...54S,2016ApJ...822...26G} and the changing mass accretion rate around the central object could possibly be accountable for this trend \citep{2011ApJ...731...50S,2014ApJ...783..105R}.\par

Clearly, the trends observed for different continuum variability parameters and spectral index changes are consistent with each other. The clear separation in the distribution of these parameters between the \emii and \dismain samples points to the significant role of ionization in BAL appearance/disappearance.

\subsection{Comparison of reddening parameters}
\label{subsec:spectral_reddening}

\begin{figure}
\includegraphics[scale=0.5]{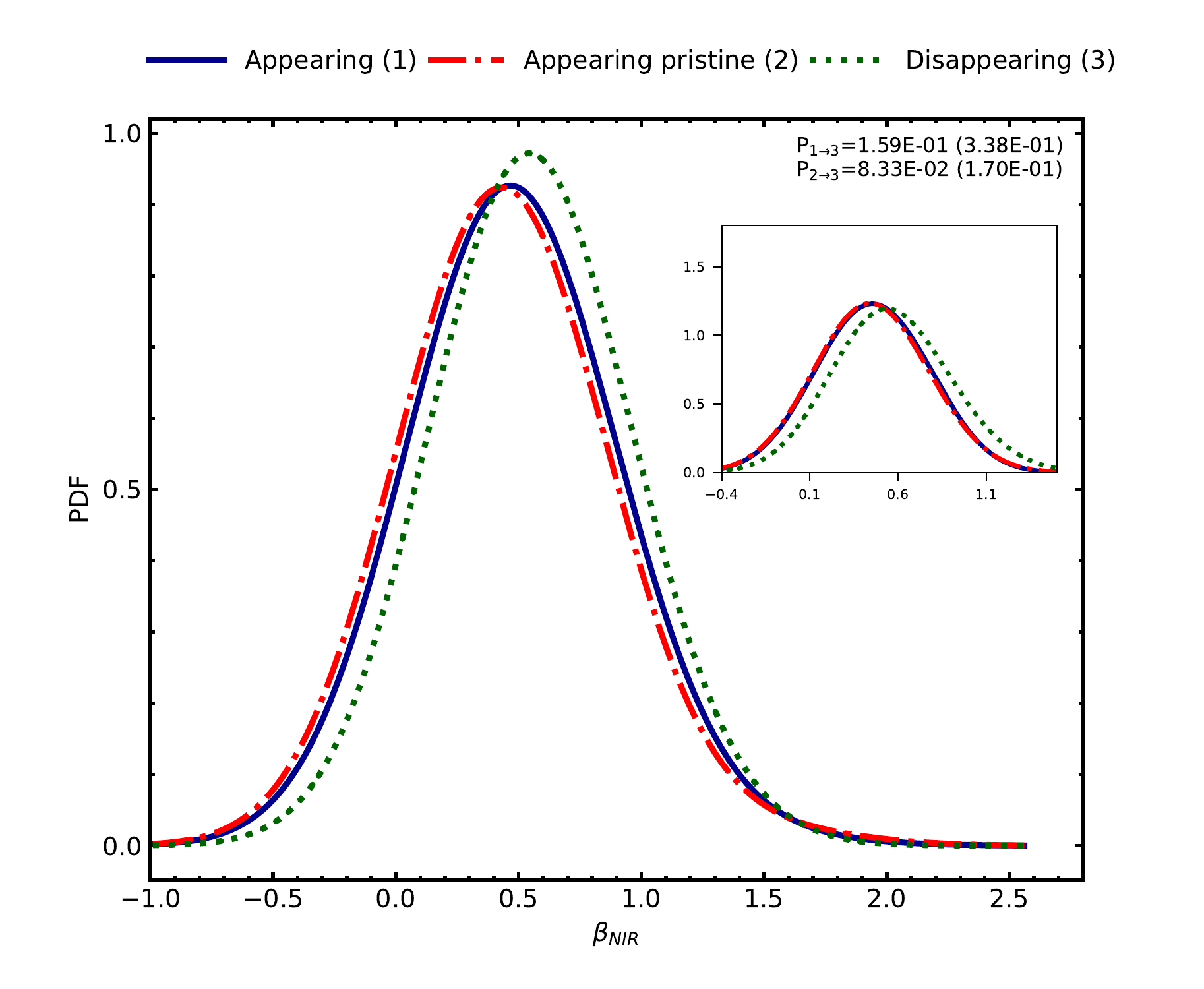}
\caption{Distribution of slope of IR continuum generated with KDE taking Gaussian kernels of similar bandwidth for \emii (\emph{solid blue}), \emiipri (\emph{dashed red}) and \dismain (\emph{dotted green}) samples. In the inset the similar distribution for the control sample of \emii, \emiipri and \dismain BAL quasars are presented. The p-values derived from the K-S test between \emii and \dismain (P$_{1 \rightarrow 3}$), and \emiipri and \dismain (P$_{2 \rightarrow 3}$) with corresponding values for the control samples are mentioned at the top right corner.}
\label{fig:red_comp}
\end{figure}

Using a sample of 2,099 SDSS BAL quasars, the study of \citet{2014ApJ...786...42Z} revealed that the outflow velocity and strength moderately correlate with the slope of the infrared continuum (\NSbetanir) which acts as a good indicator of hot dust emission. They suggested a BAL-dust scenario where dust being intrinsic to BAL outflows may contribute to the acceleration when subjected to the central radiation field. Besides, the study of \citet{2016arXiv161103733G} showed that the SED of BAL and non-BAL quasar do not intrinsically differ but apparently differ due to the dust.\par

In this section, to probe whether dust is intrinsic to the BAL outflows, we investigate the nature of variation of the infrared fluxes between the \emii and \dismain samples \citep{2016MNRAS.455..136V,2017MNRAS.467.4763T}. These sources, having undergone an extreme transition of BAL troughs, provide an ideal opportunity to verify the role of dust associated with the BAL outflows. We used the Wide-field Infrared Survey Explorer \citep[WISE;][]{2010AJ....140.1868W} catalog to search for the infrared counterparts of our BAL samples within 3$\arcsec$ from the optical position of our appearing and disappearing BAL quasars. WISE survey maps the sky in four bands W1, W2, W3, and W4 with center wavelengths at 3.4, 4.6, 12, and 22 $\mu$m respectively.  In this study, we only used reliable measurements of W1, W2, and W3 magnitudes  as the errors on W4 magnitudes for a majority of the sources were not available. The median redshifts of our BAL quasar samples are close to 2, hence covers a near-infrared (NIR) range of 1 $-$ 4$\mu$m in the rest-frame wavelengths of W1, W2, and W3 bands. For a typical quasar SED, the emission from the dust dominates beyond 3$\mu$m. To characterize the hot dust emission, several studies have used SDSS r$-$WISE W4 for high redshifts objects \citep{2016MNRAS.455..136V}. Since for the majority of quasars in our sample, reliable measurements of W4 magnitudes were not available, and W1, W2, and W3 magnitudes were not good representatives of hot dust emissions, we did not use SDSS r$-$WISE W4 to measure the excess of hot far-infrared emission relative to the UV emissions.\par

Similar to \citet{2014ApJ...786...42Z}, we used the \betanir parameter to confirm the role of dust in BAL quasars that have undergone a complete appearance or disappearance of their BAL troughs. We converted the WISE W1, W2, and W3 magnitudes to monochromatic luminosities and fitted a power law (L$_{\lambda} \propto \lambda^{\beta_{\scaleto{\rm NIR \rm}{3.5pt}}}$) in the rest-frame NIR SED of each BAL quasar in \NSemii, \NSemiipri, \NSdismain, and their corresponding control samples. The distributions \betanir are shown in Fig.~\ref{fig:red_comp} for \NSemii, \NSemiipri, and \dismain samples. It is evident from the figure that the \betanir distribution is the same for appearing and disappearing BAL quasars (similarly for their control samples) implying that dust does not have significant role in causing the extreme variations of BAL troughs. Note that \citet{2016MNRAS.455..136V} also did not find any difference in the r$-$W4 distributions of the parent radio-detected BAL sample of 60 objects compared to a sub-sample of 6 radio-detected transient BAL quasars hinting no clear role of dust in the transient BAL quasars.\par

\section{Discussion}
\label{Section:discussion}
The ionization state and covering fraction of the absorbing gas are the two parameters that determine the strength of absorption lines in quasars. Any variation in these parameters can result in changes in the absorption line profile. The BAL variability caused by changes in these two key parameters can be described by the following two scenarios: (i) change in the number of ionizing photons impinging on the absorbing clouds (IC scenario), and/or (ii) transverse motion of the gas cloud along the line of sight of the observer (TM scenario).\par  

Many previous studies on BAL variability, based on the (i) lack of correlation between the changes in ionizing continuum and absorption strength and (ii) variation in only a portion of BAL trough profile, support TM scenario \citep{2007ApJ...656...73L,2008ApJ...675..985G,2008MNRAS.391L..39H,2011MNRAS.413..908C,2013MNRAS.429.1872C,2014MNRAS.440..799V,2019ApJ...870L..25Y}. On the other hand, with an anti-correlation between the varying continuum and absorption strength, many previous studies support IC scenario \citep{1994PASP..106..548B,2012MNRAS.422.3249C,2015MNRAS.454.3962H,2015ApJ...814..150W,2017ApJS..229...22H,2018ApJ...862...22R,2018MNRAS.473L.106L,2019MNRAS.487.2818H,2019ApJ...883...30L,2019MNRAS.486.2379V}. The coordinated variation of different velocity BAL troughs of same species, or the coordinated variation of same velocity BAL troughs of different species point to the IC scenario as the main driver of BAL variations  \citep{2013ApJ...777..168F,2015ApJ...814..150W,2018ApJ...862...22R}. By definition, BAL appearance and disappearance mean a coordinated variation of BAL troughs over a range of velocities, and it is expected that the IC scenario has a significant role in driving these variations. There are also many cases where both the IC and TM scenarios are held responsible for BAL trough variations \citep{2011MNRAS.413..908C,2012MNRAS.422.3249C,2013MNRAS.429.1872C,2012ApJ...757..114F,2013ApJ...777..168F,2017MNRAS.469.3163M}.\par  

In this study, we contrasted the properties of two samples of BAL quasars that have shown either appearance or disappearance of BAL troughs between two spectroscopic observations. The three main results from our analysis are the following: (i) the \emii sample of BAL quasars is found to be more luminous as compared to the \dismain sample of BAL quasars (Fig.~\ref{fig:intrinsic_parameter}~a and d), (ii) the distributions of BAL trough depths and widths for the \emii and \dismain samples are different with the \emii BAL sample having shallower and wider BAL troughs as compared to the \dismain BAL sample (Fig.~\ref{fig:absorption_comp}~d and h), and (iii) the quasar continuum dimmed in the case of \emii sample of BAL quasars whereas it brightened in the case of \dismain sample of BAL quasars (Fig.~\ref{fig:continuum_comp}). Our study also suggests an inefficient role of dust intrinsic to the BAL clouds in causing the observed BAL transitions.\par

As the \emii\ BAL quasar sample has shorter probed timescales than the \dismain BAL quasar sample (Fig.~\ref{fig:time_zem_comp}), it is tempting to attribute the difference in the BAL depth distributions (i.e., \emii\ BAL quasar sample has shallower depths than \dismain\ BAL quasar sample) to the differences in the probed rest-frame timescales. To further probe this, we created control samples of  \NSemii, \NSemiipri, and \NSdismain\ BAL quasars based on the probed timescales and compared the BAL trough parameters. Even in these control samples based on probed timescales, we note that the distributions of |\NSdeld|, \NSavgd, and \delv are markedly different for the \emii\ and \dismain\ BAL quasar samples. Hence, the difference in BAL depth and BAL width distributions cannot be explained by the difference in the probed timescales. However, the difference in the distribution of bolometric luminosity between the \emii\ and \dismain\ BAL quasar sample may explain the observed BAL trough trends between the two samples. We did not attempt to create a control sample based on both redshift and luminosity as the sample size was significantly reduced.

The ionization parameter (log U) is the key parameter that controls the ionic response of any ion \citep{2015ApJ...814..150W}. Fig.~4 of \citet{2001ApJ...550..142H} shows the ion fraction of several ions as a function of the ionization parameter. The fraction of \civ~  increases first, then reaches a peak, and decreases after as the ionization parameter increases. Depending on the ionization parameter, the ionic response to a change in ionization parameter can be both positive and negative.  The negative ionic response (i.e., BAL disappearance accompanied by continuum brightening and vice versa) found in this study shows that the ionization parameter of these outflows is towards the right side of the peak ionization parameter (i.e., log U = -2). BAL appearance can be understood as the case where the \civ column density in the first epoch is too low to be detected (i.e., high U), but as the continuum dims (i.e., a decrease in U ), recombination of \cv to \civ increases the \civ column density resulting in the BAL appearance. Similarly, BAL disappearance occurs as a gas cloud in the low U regime is pushed to the high U regime due to an increase in the continuum. If we assume that the density of these absorbers is similar on average, our finding that the \emii BAL quasars are brighter than the \dismain BAL quasars during their earlier epochs supports the fact that the ionization parameter of the  \emii BAL quasars is higher compared to the \dismain BAL quasars before the BAL transition event.  It is known that a minimum threshold of \civ column density is required for the BAL to be detected. For a particular total hydrogen column density, Fig.~11 of \citet{2015ApJ...814..150W} shows that we can define two ionization parameters corresponding to the peak and the threshold ionic column densities. In the case of disappearing BALs, the ionization parameter in the earlier epoch can only be distributed between these two ionization parameters. However, in the case of appearing BALs, the ionization parameter in the earlier epoch can be anywhere beyond the threshold ionization parameter. If we assume that the continuum changes between the two epochs are similar for the \emii and \dismain BAL samples (a valid assumption as we have similar distributions of \NSsigdelm), the difference in the distribution of initial ionization parameters around the threshold ionization parameter may explain the difference in the distribution of \NSavgd~ and \NSdeld~ distributions.

Similarly, contrast in the distribution of the BAL widths of \emii and \dismain BAL quasars may be explained by the difference in the luminosity of these samples. Since the \emii BAL quasars are more luminous than the \dismain BAL quasars, the larger radiative forces in these sources may result in a larger velocity dispersion among the BAL cloudlets, producing a wider BAL trough.

The changing ionizing continuum has a negligible effect on the variabilities of saturated BAL troughs. The role of change in covering fraction in response to the varying BAL strength is critical in such cases \citep{2008MNRAS.391L..39H,2014MNRAS.444.1893C,2017MNRAS.469.3163M,2019MNRAS.486.2379V}. Since the average depth of our BAL troughs for both the appearing and disappearing BAL samples are less than 0.4 (see Figure~\ref{fig:absorption_comp}~d), it is reasonable to say that the effect of line saturation may be insignificant in our study \citep[see][]{2019MNRAS.486.2379V}.

If BAL variability is driven by variations in the intrinsic dust content of the absorbing gas, we expect a difference in the distribution of the hot dust emission properties between the \emii and \dismain samples. As we do not see any difference in the distribution of \betanir parameter between the two samples, we conclude that dust emission is similar for the \emii and the \dismain samples. Consequently, the dust driven BAL variability is less likely to be a favorable scenario in explaining BAL appearance/disappearance.

Conversely, it is possible that the entire BAL cloud could have moved into our line of sight between the two spectroscopic epochs (TM scenario). In such a scenario, it is impossible to explain the observed  connection between the BAL appearance/disappearance and continuum variations. Therefore, based on our findings presented in this study, we conclude that changes in the ionization state of the absorbing gas are the primary driver of extreme variations in BAL troughs.

\section{Conclusions}

Below are the point-wise conclusions of our analysis based on the three subclasses of extreme variable BAL quasars:\par

\begin{enumerate}

\item We have isolated a set of 107 appearing broad absorption line quasars based on the comparison of SDSSDR-7, SDSSDR-12, and SDSSDR-14 quasar catalogs.

\item Our \emii BAL quasars are found at relatively higher redshift with probed rest-frame time scales shorter than the \dismain BAL quasars (Fig.~\ref{fig:zem_cs} and ~\ref{fig:time_zem_comp}). In order to minimize the effect of redshift in the comparison of  \emii and \dismain BAL samples, we  also constructed the redshift-matched control samples of \NSemii, \NSemiipri, and \dismain BAL quasars. We find that the probed rest-frame timescales of the BAL appearance is shorter than the BAL disappearance even after correcting for the difference in the redshift distributions (inset of Fig.~\ref{fig:time_zem_comp}).

\item Our analysis shows that the \emii and \emiipri BAL quasars are more luminous and brighter compared to \dismain BAL quasars (Fig.~\ref{fig:intrinsic_parameter}~a and d).

\item The distributions of \NSavgew, \NSavgd, \NSdeld, and \delv indicate that the \emii BAL troughs are stronger, shallower and wider compared to the \dismain BAL troughs (see Fig.~\ref{fig:absorption_comp}). The difference in the probed timescales cannot explain the difference in the BAL trough properties. The difference in luminosity between the \emii\ and \dismain\ samples are more likely to explain the difference in BAL trough parameters.

\item We conclusively demonstrate that \emii\ and \dismain\ BAL quasars have different continuum variability trends. Based on the distributions of \NSfraccon, \NSmeddelm, and \NSslopedelm, our analysis shows the appearance of BAL troughs is accompanied by the dimming of the continuum while the disappearance of BAL troughs is accompanied by the brightening of continuum flux (see Fig.~\ref{fig:continuum_comp}~a, b, and d). The amount of continuum variation in appearance and disappearance of BAL troughs is similar as can be inferred from the similar distribution of \sigdelm in \NSemii, \NSemiipri, and \dismain samples (Fig.~\ref{fig:continuum_comp}~c).
  
\item We report a significant difference in the distribution of change in spectral index of \emii and \emiipri compared to \dismain BAL quasars, with the distribution of appearing samples (i.e. \emii and \emiipri) shifted towards the spectral softening as compared to \dismain sample which show spectral hardening (Fig.~\ref{fig:alpha_comp}). The continuum variations are in agreement with  the well known bluer-when-brighter trend seen in normal quasars, which hints to the fluctuations of the ionizing radiations are caused due to changes in the central accretion processes.

\item Our analysis also reveals that the distributions of slope of IR continuum are similar for the appearing and the disappearing samples (Fig.~\ref{fig:red_comp}), indicating BAL appearance/disappearance in the present study is less likely to be explained in terms of dust outflow scenario proposed by \citet{2014ApJ...786...42Z}.

\item Our findings support the scenario where changes in the ionizing conditions of the absorbing gas are the primary driver of  extreme variations of BAL troughs.
\end{enumerate}
 
Additional spectra for our sample of appearing BAL quasars are currently being obtained in our ongoing spectral variability campaign using 2m class telescopes. Future spectroscopic monitoring of these objects will help to put further constraints on the exact nature of variability in these sources and understand the dynamical evolution of BAL outflowing clouds.

\section*{Acknowledgments}
We thank the referee Prof. N. Filiz Ak for the feedback which has significantly helped to improve the paper.
MV acknowledges support from DST-SERB Ramanujan Fellowship (SB/S2/RJN/-018/2019).\par
Funding for SDSS-III has been provided by the Alfred P. Sloan Foundation, the Participating Institutions, the National Science Foundation, and the U.S. Department of Energy Office of Science. The SDSS-III web site is http://www.sdss3.org/. SDSS-III is managed by the Astrophysical Research Consortium for the Participating Institutions of the SDSS-III Collaboration including the University of Arizona, the Brazilian Participation Group, Brookhaven National Laboratory, Carnegie Mellon University, University of Florida, the French Participation Group, the German Participation Group, Harvard University, the Instituto de Astrofisica de Canarias, the Michigan State/Notre Dame/JINA Participation Group, Johns Hopkins University, Lawrence Berkeley National Laboratory, Max Planck Institute for Astrophysics, Max Planck Institute for Extraterrestrial Physics, New Mexico State University, New York University, Ohio State University, Pennsylvania State University, University of Portsmouth, Princeton University, the Spanish Participation Group, University of Tokyo, University of Utah, Vanderbilt University, University of Virginia, University of Washington, and Yale University.

\section*{Data availability}
The spectral data underlying this article are available in the SDSS-I/II/III Archive at https://dr14.sdss.org/optical/spectrum/search.
\bibliography{references}
\label{lastpage}
\end{document}